# What do we know about the disruption index in scientometrics?
# An overview of the literature

Christian Leibel*+ & Lutz Bornmann*

*Science Policy and Strategy Department

Administrative Headquarters of the Max Planck Society

Hofgartenstr. 8,

80539 Munich, Germany.

Email: bornmann@gv.mpg.de; christian.leibel.extern@gv.mpg.de (corresponding author)

ORCID ID: 0000-0003-0810-7091 (LB); 0009-0006-6893-4901 (CL)

+Ludwig-Maximilians-Universität München

Department of Sociology

Konradstr. 6,

80801 Munich, Germany.

# Abstract

The purpose of this paper is to provide a review of the literature on the original disruption index ($DI_1$) and its variants in scientometrics. The $DI_1$ has received much media attention and prompted a public debate about science policy implications, since a study published in *Nature* found that papers in all disciplines and patents are becoming less disruptive over time. This review explains in the first part the $DI_1$ and its variants in detail by examining their technical and theoretical properties. The remaining parts of the review are devoted to studies that examine the validity and the limitations of the indices. Particular focus is placed on (1) possible biases that affect disruption indices (2) the convergent and predictive validity of disruption scores, and (3) the comparative performance of the $DI_1$ and its variants. The review shows that, while the literature on convergent validity is not entirely conclusive, it is clear that some modified index variants, in particular $DI_5$, show higher degrees of convergent validity than $DI_1$. The literature draws attention to the fact that (some) disruption indices suffer from inconsistency, time-sensitive biases, and several data-induced biases. The limitations of disruption indices are highlighted and best practice guidelines are provided. The review encourages users of the index to inform about the variety of $DI_1$ variants and to apply the most appropriate variant. More research on the validity of disruption scores as well as a more precise understanding of disruption as a theoretical construct is needed before the indices can be used in the research evaluation practice.

**Key words**

scientometrics, bibliometrics, disruption index, CD index, disruption index

# 1 Introduction

Only five years have passed since the introduction of the disruption index (DI$_1$) by Funk and Owen-Smith (2017)[1], and meanwhile it has seen widespread application. Many researchers have used the DI$_1$ to identify the most disruptive publications in specific disciplines and/or subdisciplines. Numerous articles, especially in the field of life sciences, have applied the DI$_1$ to the field-specific literature to identify disruptive publications in different disciplines: surgery (Becerra et al., 2021; Becerra et al., 2022; Hansdorfer et al., 2021; Horen et al., 2021; Sullivan et al., 2021; Williams et al., 2021), radiology (Abu-Omar et al., 2022), breast cancer research (Grunvald et al., 2021), urology (Khusid et al., 2021), ophthalmology (Patel et al., 2022), energy security (Jiang & Liu, 2023a), and nanoscience (Kong et al., 2023). In the field of scientometrics, Bornmann and Tekles (2019b), and Bornmann et al. (2020b) tried to find the most disruptive papers published in *Scientometrics* with the help of (a modified version of) the DI$_1$. The popularity of the new index is not only reflected in its application in several disciplines, but also in the recent introduction of an index variant on the journal level. Jiang and Liu (2023b) proposed the Journal Disruption Index (JDI) as an alternative to (traditional) journal level metrics such as the Journal Impact Factor (JIF, provided by Clarivate). Furthermore, Yang, Hu, et al. (2023) and R. Wang et al. (2023) proposed different ways to incorporate the DI$_1$ to the evaluation of scientists' research impact.

The DI$_1$ played a key role in two influential science of science papers published recently in *Nature*: (1) Wu et al. (2019) used the DI$_1$ to investigate how the growth of team science impacts research outputs. They found that large teams tend to conduct consolidating research while small teams tend to produce disruptive publications. Although (international) cooperation is frequently seen as key factor for scientific excellence, disruptive research seems to be connected with rather small research groups. (2) Park et al. (2023) shocked the scientific community (and beyond) with the claim that scientific papers and patents have been getting less disruptive since World War II. Using data on 45 million papers and 3.9 million patents, they report that there has been a continuous decrease in average disruption scores across all disciplines. The article made waves in and beyond the science system and prompted a public debate surrounding the question of if and why science is running out of steam in spite of the massive expansion of the (global) science system in recent decades.

---

[1] The authors called the disruption index CD index.

While the finding that both patents and papers are getting »less bang per buck« is certainly spectacular, it is important not to jump unreflectively and straight forward to far reaching conclusions (science policy actions). Park et al. (2023, p. 143) themselves point out that "even though research to date supports the validity of the CD index [referred to as $DI_1$ in this review], it is a relatively new index of innovative activity and will benefit from future work on its behaviour and properties". Therefore, any meaningful discussion about the results of Park et al. (2023) (as well as the results of any other study involving the $DI_1$) requires a detailed understanding of the index's properties and limitations, which have been studied in several (empirical) studies since 2019.

In order to provide detailed insights into the properties and limitations of (different variants of) the $DI_1$, this review paper provides a systematic review of the current literature on $DI_1$ and its modified variants. The review consists of three parts. In the first part, the technical and theoretical properties of the $DI_1$ are explained. The second part covers the numerous modified index variants of the $DI_1$ researchers have proposed so far. The third part provides an overview of the literature on the validity of disruption scores. This part discusses the studies dealing with the important question whether the indices measure what they propose to measure.

## 2   Definition and history of the disruption index

### 2.1   Creation and calculation of the disruption index

The $DI_1$ was created by Funk and Owen-Smith (2017) in order to quantify the magnitude of technological change brought about by new patents.[2] Long before the creation of the new index, researchers had already observed that there are two distinct types of technological shifts: "[M]ajor technological shifts can be classified as *competence-destroying* or *competence-enhancing* [emphasis in original] [...], because they either destroy or enhance the competence of existing firms in an industry. The former require new skills, abilities, and knowledge in both the development and production of the product. The hallmark of competence-destroying discontinuities is that mastery of the new technology fundamentally

---

[2] Funk and Owen-Smith (2017) created the $DI_1$, but the idea of using citation data to identify transformative research was proposed in earlier publications. For example, Huang et al. (2013, p. 291) stated in a conference paper: "We view the process by which transformative research is recognized by the scientific community as a competition between paradigms for the attention of the scientific community. [...] We claim that transformative research shifts attention of the scientific community away from the established paradigm and that this is observable as a disruption of the growth of its citations cascade. Disruption occurs when the challenger paradigm can explain new citations received by the established paradigm".

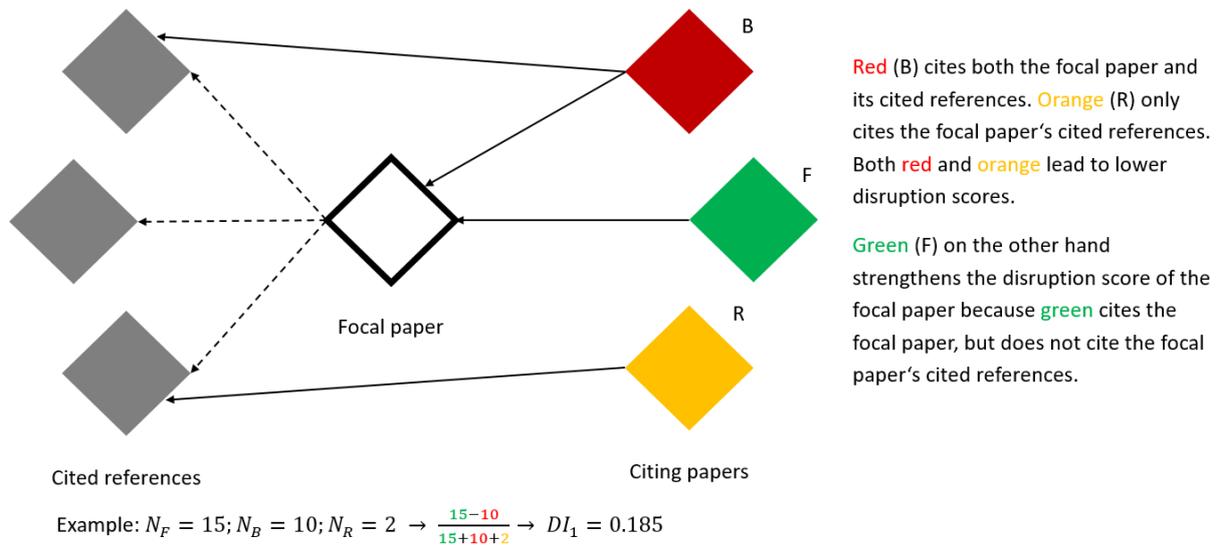

Red (B) cites both the focal paper and its cited references. Orange (R) only cites the focal paper's cited references. Both red and orange lead to lower disruption scores.

Green (F) on the other hand strengthens the disruption score of the focal paper because green cites the focal paper, but does not cite the focal paper's cited references.

Focal paper

Cited references

Citing papers

Example: $N_F = 15; N_B = 10; N_R = 2 \rightarrow \frac{15-10}{15+10+2} \rightarrow DI_1 = 0.185$

*Figure 1: Calculation of the DI₁ in a tripartite network. The illustration is based on Funk and Owen-Smith (2017).*

alters the set of relevant competences within a product class" (Tushman & Anderson, 1986, p. 442). In other words, some technological innovations improve upon established technologies without replacing them, whereas others render previous technologies obsolete. Funk and Owen-Smith (2017) took manifold inspiration from the literature on technological shifts, but they were of the opinion that the dichotomy of competence-destroying or competence-enhancing technologies lacked nuance. They argued that "a new technology's influence on the status quo is a matter of degree, not categorical influence" (Funk & Owen-Smith, 2017, p. 792). Funk and Owen-Smith (2017) also claimed that established measures of technological impact (like citation counts) only capture the magnitude of a technology's use and thus miss "the key substantive distinction between new things that are valuable because they reinforce the status quo and new things that are valuable because they challenge the existing order" (Funk & Owen-Smith, 2017, p. 793). Therefore, they created the DI₁ that could take advantage of vast patent databases like the U.S. Patent Citations Data File. Since innovation is a valuable resource not just in the realm of technology (measured by patents and their citations data), but also in the realm of science (measured by publications and their citation data), the concept of disruption attracted the attention of Wu et al. (2019), who were the first to apply the DI₁ to the world of bibliometrics.

The DI₁ is based on citation networks (Figure 1). Each citation network consists of three elements: a focal paper (FP), a set of references cited by the FP (set R) and a set of citing papers (set C). The citing papers are divided into three mutually exclusive groups. Group F (for "FP") encompasses all publications that cite the FP without citing even a single one of the FP's cited references. Publications that cite both the FP and at least one of its cited references belong in group B (for "both"), whereas group R (for "reference") consists of publications that cite at least one of the FP's cited references without citing the FP itself. $N_F$, $N_B$ and $N_R$ represent the total number of papers in set F, B, and R, respectively.

The interpretation of $N_F$ and $N_B$ is rather straightforward: A large $N_F$ indicates that the FP renders its own cited references obsolete and is thus associated with highly disruptive publications. In contrast, a large $N_B$ is a sign of a consolidating publication because the citation impact of the FP is dependent on the citation impact of its references. The intended purpose of $N_R$ is to weaken the disruption value of the FP, but this only works if the numerator ($N_F - N_B$) is positive. However, in case of ($N_F - N_B$) < 0, $N_R$ actually strengthens the disruption score of the FP (in the sense of being less consolidating). Since this inconsistency poses a significant thread to the validity of disruption scores, more information on this topic will be presented in Section 4.1. The $DI_1$ is equivalent to the following ratio:

$$DI_1 = \frac{N_F - N_B}{N_F + N_B + N_R}$$

The $DI_1$ has a range of -1 to 1. Negative values are supposed to indicate developmental papers whereas positive values supposedly signify disruptive papers. Two things should be kept in mind about the calculation of the $DI_1$: First, the $DI_1$ is based on bibliographic coupling. Bibliographic coupling is a method that looks for publications that cite the same references. The $DI_1$ applies bibliographic coupling to FPs and their citing papers. Consequently, one might argue that the $DI_1$ "can be considered as a continuity indicator more than a disruption indicator since the operation is grounded in bibliographic coupling. The bibliographic coupling of a focal paper to its references generates a representation of continuity. From this perspective, discontinuity is indicated when the bibliographic coupling is not sufficiently generating continuity" (Leydesdorff et al., 2021).

This point ties in with a second point: Following the terminology proposed by Bu et al. (2021), $DI_1$ is a *relative* index because it treats disruption and consolidation as opposite concepts. From an absolute perspective, the citation network of an FP may simultaneously contain many bibliographic couplings links (indicating consolidating science) and a large $N_F$ (indicating disruptive science). In absolute terms, such an FP is both highly disruptive and highly consolidating. In contrast, from a relative perspective, the relationship between disruption and consolidation is a zero-sum game: No publication may be both disruptive and consolidating at the same time. For example, an article with a $DI_1$ score of 0.5 is supposed to be more disruptive and less consolidating than an article with a $DI_1$ score of 0. An article with a $DI_1$ score of 0.3 is less disruptive and more consolidating than an article with a $DI_1$ score of 0.4.

## 2.2 The disruption index's underlying theoretical concepts

In this section, implicit theoretical assumptions built into the $DI_1$ (and its modified variants) are explained in relation to two important theoretical concepts: the concept of novelty and the concept of scientific revolutions. Although the literature does not provide a precise definition of the term »disruption«, it can be said with certainty that there are significant differences between the concept of »disruption« and the concept of »novelty«. Research on novelty indices predates the creation of the $DI_1$ by a couple of years (Foster et al., 2015; Lee et al., 2015; Uzzi et al., 2013). Novelty indices are guided by the notion that creativity is no *creatio ex nihilo* but rather a cumulative process that manifests in atypical combinations of prior knowledge. According to Lee et al. (2015, p. 685), novelty indices were born out of a stream of research that "views creativity as an evolutionary search process across a combinatorial space and sees creativity as the novel recombination of elements". For example, researchers calculated the novelty value of papers by searching their bibliography for atypical (Uzzi et al., 2013) or unique (Wang et al., 2017) combinations of cited references. In contrast to novelty indices, which only consider the cited references of an FP, the $DI_1$ also considers the FP's citing papers. This is not just a technical, but also a conceptual difference. Novelty indices focus on the origin of creative ideas in combinatorial processes. But, as Lee et al. (2015) explain, creativity is not just about the origin of ideas, it is also their usefulness and impact that matters. By also considering citing papers in its calculation, the $DI_1$ captures not just the origin, but also the impact of new ideas. This is intuitively plausible since a novel idea that receives barely any attention from the scientific community hardly deserves to be labelled »disruptive«: "Although novelty may be necessary for disruptiveness, it is not necessarily sufficient to make something disruptive" (Bornmann et al., 2020a, p. 1256).

The conceptual difference between disruption and novelty is also reflected in empirical results. By examining a dataset on Citation Classics, Leahey et al. (2023) showed that only specific types of novelty are linked to higher disruption scores. In the Citation Classics dataset, new methods are positively associated with disruption scores, whereas new theories and new results are negatively associated with disruption scores. Shibayama and Wang (2020) investigated the relationship between two types of novelty (theoretical and methodological) and disruption scores (see Section 5.4). The study is based on data from a survey, which asked researchers to rate the theoretical and methodological originality of their own publications. Shibayama and Wang (2020) found that disruption scores are positively associated with self-assessed theoretical originality, but not with self-assessed methodological originality. Even though it is difficult to draw conclusions from two studies that employed different methods and produced seemingly contradictory results, both Shibayama and Wang (2020) and Leahey et al. (2023) highlight that only a specific subset of novel research is also disruptive research.

In addition to novelty, the $DI_1$ also relies heavily on concepts inspired by Kuhn's theory of paradigm shifts (Kuhn, 1962). According to Kuhn, the history of science can be categorized into two repeating phases: normal science and scientific revolutions. Normal science is characterized by the modus operandi of a specific paradigm: "For Kuhn science progresses by gradual, incremental changes in a particular discipline's practice and knowledge" (Marcum, 2015, p. 143). The phase of normal science is brought to an end by sudden paradigm shifts caused by scientific breakthroughs that drastically alter the status quo. Negative (or low) disruption scores are often interpreted as representations of the consolidating nature of normal science, whereas positive (or high) disruption scores are supposed to indicate drastic scientific breakthroughs or even paradigm shifts (Bornmann et al., 2020a; Bornmann & Tekles, 2021; Li & Chen, 2022; Liang et al., 2022; Shibayama & Wang, 2020; S. Wang et al., 2023).

# 3   Variants of the disruption index

Since the introduction of $DI_1$ by Funk and Owen-Smith (2017), a number of researchers have suggested modified variants of the index. These variants will be explained in this section.[3] The explanations do not follow a chronological order; instead the different variants are categorized into distinct groups based on their specific type of modification.

## 3.1   Disruption and citation impact

The first alternative to $DI_1$ was suggested by Funk and Owen-Smith (2017) themselves. In addition to $DI_1$, they also constructed $mDI_1$. The difference between the two indices is the inclusion of the weighting parameter $m_t$, which captures only those citations directly linked to the FP.

---

[3] Jiang and Liu (2023b) mention three variants of the $DI_1$ that we are unfortunately not able to cover in this review because the papers they were proposed in are only available in Chinese. These two papers are Liu et al. (2020) and Song et al. (2022).

$$mDI_1 = \frac{m_t}{n_t} \times \frac{N_F - N_B}{N_F + N_B + N_R}$$

"In this formulation, $m_t$ differs from $n_t$ in that the former counts only citations of the focal patent, whereas the latter includes citations of both the focal patent and its predecessors" (Funk & Owen-Smith, 2017, p. 795). Whereas $DI_1$ "does not discriminate among inventions that influence a large stream of subsequent work and those that shape the attention of a smaller number of later inventors" (Funk & Owen-Smith, 2017, p. 795), $mDI_1$ also accounts for the magnitude of a patent's use. Even though $mDI_1$ so far has received little attention from researchers, the idea of an index that measures both citation impact and disruption is not without merit.

Consider the hypothetical case of two papers A and B: A and B are assigned identical $DI_1$ scores, but A's citation impact by far surpasses B's citation impact. This in turn means that A inspired many researchers to pursue new ideas, whereas B did not have a lasting impact on the scientific community. While there are good reasons to differentiate between low and high impact research, Wei et al. (2023) argue that the influence of citation impact is too dominant in the calculation of $mDI_1$ because of the different scaling of citation counts and disruption scores.

As an alternative to distilling citation impact and disruption values down to one number, Wei et al. (2023) constructed a two-dimensional framework that keeps the measurement of citation impact and disruption separate (Figure 2). In this framework, publications with both high citation counts and high disruption scores are classified as revolutionary science. Articles like paper B in the example above fall in the low impact direction-changing science category because they introduce original ideas but do not gain the recognition of many researchers. High impact incremental science represents influential consolidating research. Most articles are low impact incremental science since they neither contain revolutionary ideas nor do they gain a lot of attention in the form of citations.

Wei et al. (2023) drew the line between consolidating and disruptive science at a $DI_1$ value of 0, taking advantage of the fact that negative $DI_1$ values are supposed to indicate consolidating publications. For the x-axis in Figure 2, the choice for the dividing line is less clear. Wei et al.

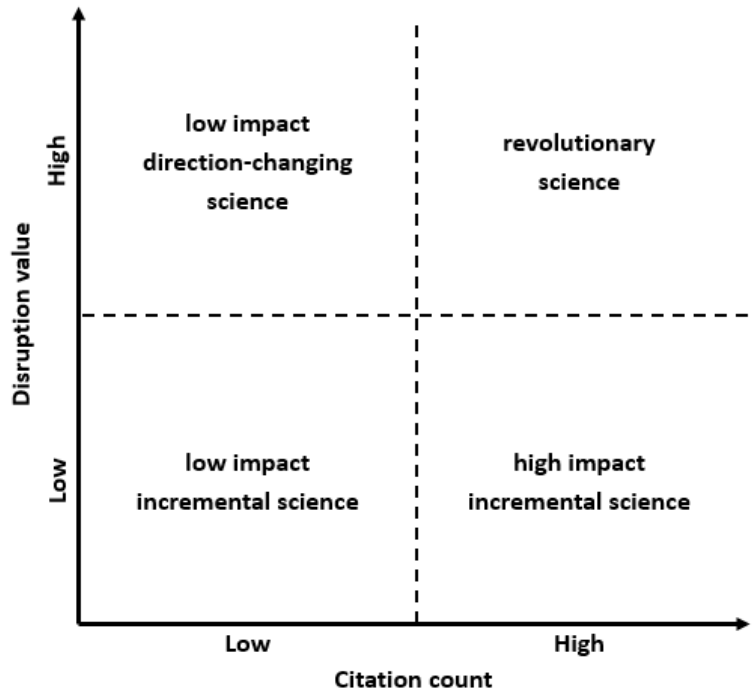

*Figure 2: Classification of consolidating and disruptive science based on Wei et al. (2023).*

(2023) used logarithmized citation counts and placed the dividing line between high and low impact science at a value of 2.0. As an alternative to logarithmized citation counts one could use the average or median citation counts (or a relative measure of citation impact like citation percentiles).

Depending on the research evaluation context, it might be worth considering not only the magnitude, but also the field-specificity of a publication's citation impact. Hypothetically, two papers A and B may have identical citation counts and disruption values, but differ greatly in the way they exert influence on the scientific community: While Paper A is a source of inspiration for scientists from many different disciplines, Paper B mainly grabs the attention of scientists working within a specific field. Since the $DI_1$ considers all citations of the FP regardless of the disciplines the citing papers belong to, it would not be able to distinguish between papers with a broad citation impact (like Paper A) and papers with a field-specific citation impact (like Paper B).[4] This is an issue if one seeks to find the most disruptive publications in a particular discipline. Therefore, Bornmann et al. (2020b) suggested an improved field-specific variant of the $DI_1$. In order to find disruptive papers published in *Scientometrics,* they redefined $N_B$ and $N_R$ as follows:

---

[4] The distinction used here is related, but not identical to the distinction between broad and deep citation impact that was introduced by Bu et al. (2021).

- $N_B^I$: Number of papers citing the FP, and at least $I$ of the cited references of all *Scientometrics* papers published in the same year as the FP.
- $N_R$: Number of papers citing at least one of the cited references of all *Scientometrics* papers published in the same year as the FP, but not the FP itself.

The reasoning behind the addition of the threshold $I$ to $N_B$ will be discussed in the next section. Following Bittmann et al. (2022), the field-specific versions of the $DI_1$ will be referred to as $DI_{1n}$. Compared to $DI_1$, $DI_{1n}$ is based on a larger set of cited references as it does not only consider the cited references of the FP, but all cited references of all papers published in a certain journal within a certain time window.

## 3.2 Dealing with noise caused by highly cited references

Recall that $N_R$ denotes the number of publications that cite at least one of the FP's cited references, but do not cite the FP itself. Since $N_R$ is part of the denominator, a large $N_R$ pushes $DI_1$ scores closer to zero (see Section 4.1). Because $N_R$ essentially captures the citation impact of the FP's cited references within the citation network, the FP's disruption value could be biased by the number of references it cites and by the citation impact of these references. Bornmann and Tekles (2021) explain this problem in detail: "Suppose that a focal paper cites a few highly cited papers, which are very likely to be cited by papers citing the focal paper, even if the focal paper is rather disruptive. In such a situation, the citing papers with only a few citation links to the focal paper's cited references may not be adequate indices for disruptive research" (see Section 4.3).

Bornmann et al. (2020a) were the first to suggest a way to eliminate biases caused by highly cited references. They modified $DI_1$ by implementing a threshold $I$ ($I > 1$), such that only those citing papers that cite at least $I$ of the FP's cited references are considered in the calculation of the index values. Whereas the $DI_1$ only takes into account whether or not there is at least one bibliographic coupling link between the FP and its citing papers, $DI_I$ also considers the strength of the bibliographic coupling links. More specifically, Bornmann et al. (2020a) recommend a threshold of $I = 5$. $DI_5$ excludes all citing papers that cite less than five of the FP's cited references and thereby focuses on citing papers that rely more heavily on the FP's cited references. In the hypothetical case of an FP that cites three highly influential publications, $DI_5$ would not consider citing papers that cite only these three publications and none of the other references cited by the FP.

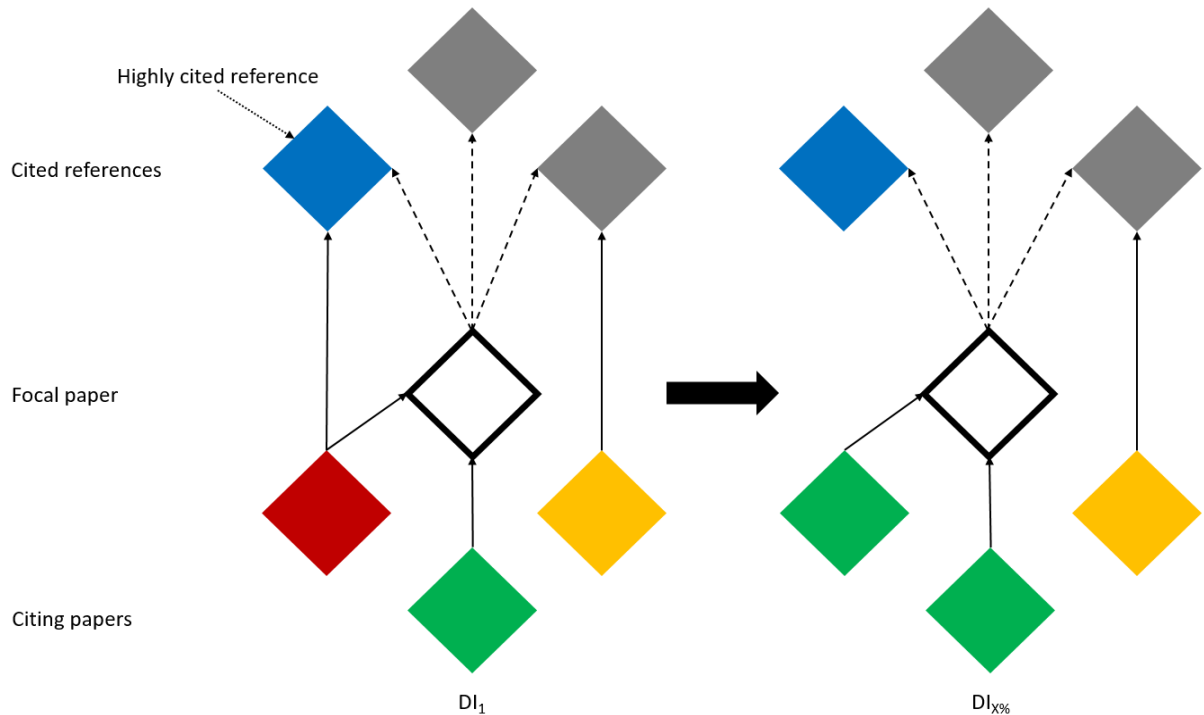

*Figure 3: Comparison of how $DI_1$ and $DI_{X\%}$ handle highly cited references. Illustration based on Deng and Zeng (2023). The colour green represents $N_F$, red represents $N_B$, and orange represents $N_R$.*

Recently, Deng and Zeng (2023) suggested a different way to get rid of the noise brought about by highly cited references. Instead of excluding citing papers that do not reach a minimum threshold of bibliographic coupling links with the FP, they opted to use a threshold $X$ such that the $X\%$ most highly cited references are selected and excluded. As Figure 3 illustrates, the exclusion of highly cited references could turn some red ($N_B$) or orange ($N_R$) citing papers into green citing papers ($N_F$). Deng and Zeng (2023) chose to refer to their new index by the simple name of "new disruption". To fit in it with the denotation used for other variants, the *new disruption* will be denoted as $DI_{X\%}$ for a threshold of $X$ (e.g. $DI_{1\%}$, $DI_{5\%}$, $DI_{10\%}$, etc.).

## 3.3   Variants without $N_R$

$DI_I$ and $DI_{X\%}$ keep $N_R$, but try to eliminate some of the noise caused by highly cited references. A potential disadvantage of indices like $DI_I$ and $DI_{X\%}$ is that they rely on arbitrary thresholds ($I = 5$ and $X = 3$). Instead of using thresholds, one could also drop $N_R$ entirely. Dropping $N_R$ results in indices considering only such citing papers that cite the FP. Wu and Yan (2019) discussed an index that corresponds to DI₁ but drops $N_R$. In line with the denotation used by Bornmann et al. (2020a), indices of this type will be referred to as DI^noR.

$$DI^{noR} = \frac{N_F - N_B}{N_F + N_B}$$

Another approach to get rid of $N_R$ was suggested by Bu et al. (2021) in a paper that introduced the dependency index (DEP).[5] The "DEP is defined as the average number of citation links from a paper citing the FP to the FP's cited references. A high (average) number of such citation links indicates a high dependency of citing papers on earlier work so that disruptiveness is represented by small values of DEP" (Bittmann et al., 2022, p. 1250).

$$DEP = \frac{T_R}{C}$$

In this formulation, $T_R$ represents the total number of bibliographic coupling links between the FP and its citing papers. $C$ is the total number of citing papers. As the name suggests, the DEP measures how strongly the citation impact of the FP depends on the citation impact of its references. Unlike the other variants discussed so far, the DEP does not have a theoretical upper bound. Because the DEP measures the opposite of disruption, low DEP values correspond to high values of other variants. An inverse version of the DEP is perhaps easier to interpret (when other index variants are also used in a study). Bittmann et al. (2022) constructed the inverse DEP by subtracting the DEP values of every FP from the empirical maximum value observed in the sample and adding 1 to the result. The inverse DEP has a theoretical upper bound of 1, but no theoretical lower bound.

A third variant of the DI₁ without $N_R$ is the Shibayama-Wang originality, named after its creators Shibayama and Wang (2020). They took advantage of the fact that dropping $N_R$ allows them to construct an index that counts the actual bibliographic coupling links instead of counting the linked publications. The originality index, denoted as $Orig_{base}$, is calculated as follows:

---

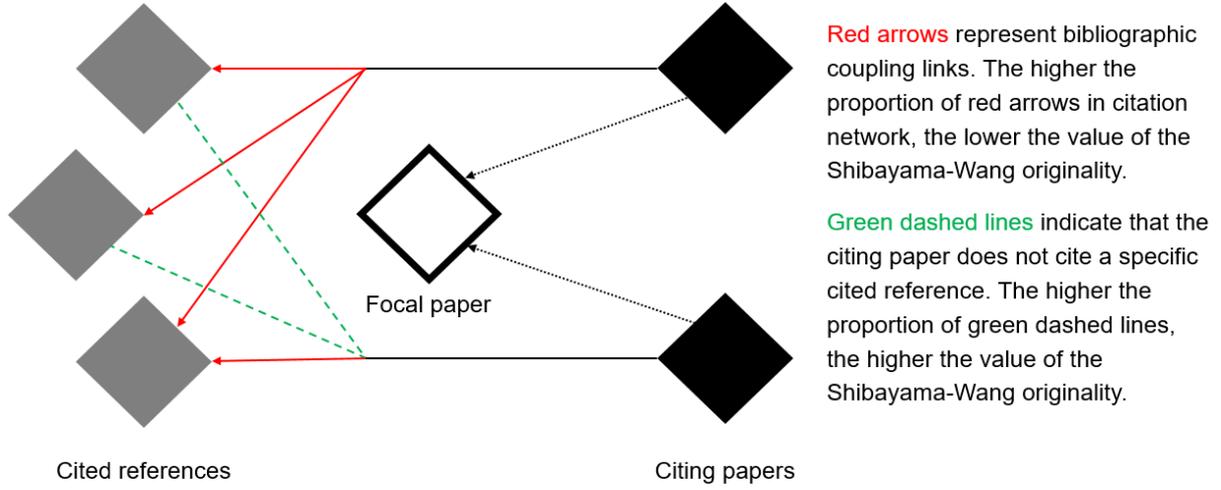

$$Orig_{base} = 1 - \frac{1}{CR} \sum_{c=1}^{C} \sum_{r=1}^{R} x_{cr} \; with \; x_{cr} = \begin{cases} 1 \; if \; c \; cites \; r \\ 0 \; otherwise \end{cases}$$

Red arrows represent bibliographic coupling links. The higher the proportion of red arrows in citation network, the lower the value of the Shibayama-Wang originality.

Green dashed lines indicate that the citing paper does not cite a specific cited reference. The higher the proportion of green dashed lines, the higher the value of the Shibayama-Wang originality.

Cited references

Citing papers

Focal paper

$$Orig_{base} = 1 - \frac{(4 \times 1) + (2 \times 0)}{2 \times 3} = 0.33$$

*Figure 4: Graphical representation of the Shibayama-Wang originality in a simple citation network. The links connecting the FP to its cited references were left out for aesthetic reasons.*

In the formula, $C$ denotes the total number of the FP's citing papers and $R$ denotes the total number of the FP's cited references. Analogously, $r$ and $c$ refer to a specific citing paper and a specific cited reference, respectively. Note that the originality index does not consider publications that only cite the FP's cited references, but do not cite the FP itself. The originality score ranges from 0 to 1 and is equivalent to the proportion of $x_{cr} = 0$ in the citation network (represented by green dashed lines in Figure 4). Like other index variants, the Shibayama-Wang originality is influenced by cited references with high citations counts. Shibayama and Wang (2020) also address the possibility that the number of cited references of the FP's citing papers could bias the calculation of the originality index because papers with many references in set $C$ are more likely to cite papers in set $R$. To tackle these two sources of bias, Shibayama and Wang (2020) suggested two weighted versions of $Orig_{base}$:

$$Orig_{weighted\_yc} = 1 - \frac{L}{R} \times \frac{\sum_{c=1}^{C} \sum_{r=1}^{R} x_{cr}}{\sum_{c=1}^{C} y_c}$$

$$Orig_{weighted\_zr} = 1 - L \times \frac{\sum_{c=1}^{C} \sum_{r=1}^{R} x_{cr}}{\sum_{c=1}^{C} y_c \sum_{r=1}^{R} z_r}$$

In the formulas, $y_c$ denotes the reference count of the $c^{th}$ citing paper; $z_r$ is the citation count of the $r^{th}$ cited reference. $L$ is an arbitrary positive number which may be chosen such that the minimum originality value equals zero.

## 3.4   Disentangling disruption and consolidation

The index variants discussed so far treat the relationship between disruption and consolidation as a trade-off, because they distil the disruptive and consolidating aspects of a given publication down to a single number. In certain cases, it may be more useful to treat disruption and consolidation not as opposites, but as two distinct concepts which require two distinct indices. As demonstrated by Leydesdorff et al. (2021), the simplest way to construct indices that serve this purpose is to change the numerator in the calculation of the $DI_1$:

$$DI^* = \frac{N_F}{N_F + N_B + N_R}; \quad DI^\# = \frac{N_B}{N_F + N_B + N_R}$$

These modified variants of $DI_1$ separate the concepts of disruption and consolidation: $DI^*$ measures disruption, whereas $DI^\#$ measures consolidation. Both indices are positive by definition and thus have a range of 0 to 1. Leydesdorff et al. (2021) illustrated the advantage of having two indices using the example of two papers, Paper A and Paper B.

- Paper A: $N_F = 10, N_B = 10, N_R = 100$
- Paper B: $N_F = 10, N_B = 100, N_R = 10$

The $DI_1$ assigns the value of 0 to Paper A and -0.75 to Paper B respectively. This might lead to the conclusion that Paper B is less disruptive. However, a more detailed inspection using $DI^*$ and $DI^\#$ reveals that $DI^*$ – focusing on disruption – assigns the same value (0.083) to both papers, implying that they are equally disruptive. The two publications only differ with respect to their consolidation values. The $DI^\#$ value is ten times larger for Paper B (0.83) than for Paper A (0.083), meaning that Paper B is more consolidating than Paper A. In addition to this example, Leydesdorff et al. (2021) also provided another more conceptual argument for the use of $DI^*$ and $DI^\#$ that relates to the weight given to $N_B$ in the calculation of the index values: "The difference between the total number of citing papers $(N_F + N_B)$ and the value in the numerator [...] is $(N_F + N_B) - (N_F - N_B) = 2 \times N_B$. One could argue that it would be more parsimonious to subtract $N_B$ only once from the total citations $(N_F + N_B)$" (Leydesdorff et al.,

2021).[6]

A different line of argument was put forward by Chen et al. (2021) in the context of research on patent data. They criticized the dichotomous typology of either competency-destroying or competency-enhancing technologies, which is fundamental to the construction of the $DI_1$, as being too one-sided. The main reason for this criticism is the failure of the dichotomous typology to identify "dual technologies" (Chen et al., 2021). Dual technologies consolidate some of their prior arts while simultaneously disrupting others: "For example, digital photography was built on electrical technology and simultaneously destabilized chemical photography" (Chen et al., 2021). Chen et al. (2021) operationalized the dual view of technology by modifying the tripartite network structure of Funk and Owen-Smith (2017), such that for every FP there is a set $p$ of prior arts (denoted as $p = [p_1, \ldots, p_i]$). $DI_1$ is split into two distinct indices D and C.[7] The calculation of D and C involves two steps: In the first step, for every prior art $p_i$ the respective $D_i$ and $C_i$ values are calculated as follows:

$$D_i = \frac{N_F^i}{N_F^i + N_B^i + N_P^i}; \quad C_i = \frac{N_B^i}{N_F^i + N_B^i + N_P^i}$$

$$D = \frac{1}{n} \times \sum_{i=1}^{n} D_i; \quad C = \frac{1}{n} \times \sum_{i=1}^{n} C_i$$

Analogous to the calculation of $DI_1$, $N_F^i$ denotes the total number of publications that cite the FP but not $p_i$, $N_B^i$ represents the total number of publications that cite both the FP and $p_i$ and $N_P^i$ is the total number of publications that cite $p_i$, but do not cite the FP. In the second step, the final D and C values are calculated by averaging across all $D_i$ and $C_i$. The D and C indices provide detailed insights into the citation networks of patents and papers. Not only do they allow for the separate calculation of disruption and consolidation values, but the respective $D_i$ and $C_i$ values also provide information about the relationship between an FP and its prior arts. Chen et al. (2021) illustrated the advantage of using separate indices for disruption and consolidation scores by constructing a more nuanced framework of technological innovation. As shown in Figure 5, dual technologies are characterized by both high D and high C values. Both this framework and the D and C indices may be repurposed for bibliometrics by simply replacing the prior arts $p_1, \dots, p_i$ with cited references $r_1, \dots, r_i$ (Li & Chen, 2022).

---

[6] Yang, Deng, et al. (2023) proposed a 2-step variant of DI* called 2stepD, which considers two generations of citing papers (i.e. papers that cite the FP's citing papers). According to Yang, Deng, et al. (2023) the 2stepD is aimed at the identification of critical nodes in networks (e.g. social networks, transportation networks and biological networks). Since there is no clear connection between the 2stepD and scientometrics, it is not covered in detail in this review.

[7] The D and C indices were first proposed in a conference paper (Li & Chen, 2017). In this review, we focus on the more fleshed out article (Chen et al., 2021).

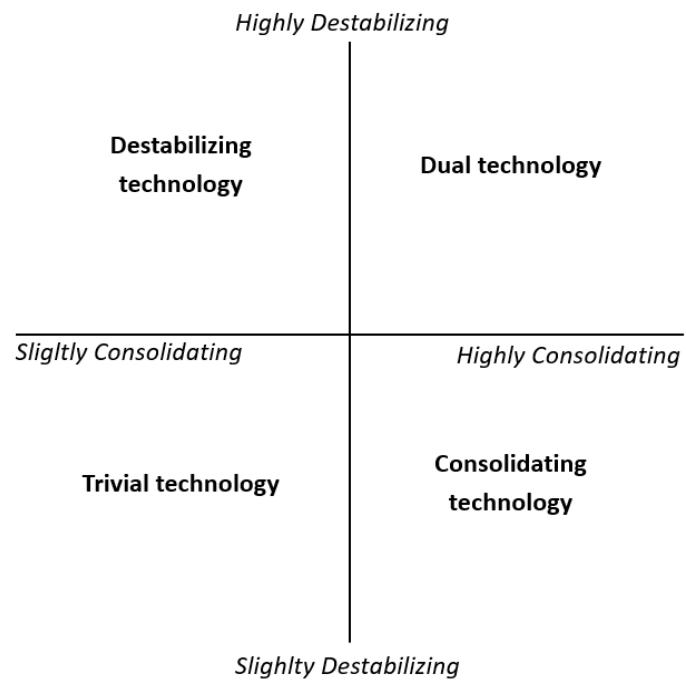

*Figure 5: Framework for the classification of consolidating and destabilizing technologies. The illustration is based on Chen et al. (2021).*

## 3.5   Measuring disruption with keywords and MeSH terms

While all studies mentioned so far try to measure disruption using citation networks, researchers have also made efforts to measure disruption and/or novelty with key words and text data (Arts et al., 2021; Boudreau et al., 2016; Foster et al., 2015; Hou et al., 2022). S. Wang et al. (2023) introduced the "Entity-based Disruption Index" (ED) in an effort to combine network-based and text-based approaches. Instead of counting citations, ED relies on keywords to capture the flow of knowledge elements within a citation network. The authors operationalized knowledge elements using Medical Subject Headings (MeSH, National Library of Medicine). MeSH terms are manually assigned by experts (in the corresponding fields) to describe the content of biomedical literature. There are two types of MeSH terms: Major topic MeSH terms, on the one hand, describe the main concepts of a publication. Subheading MeSH terms, on the other hand, provide supplementary information about a publication's content. Note that the ED does not distinguish between major topic and subheading MeSH terms.

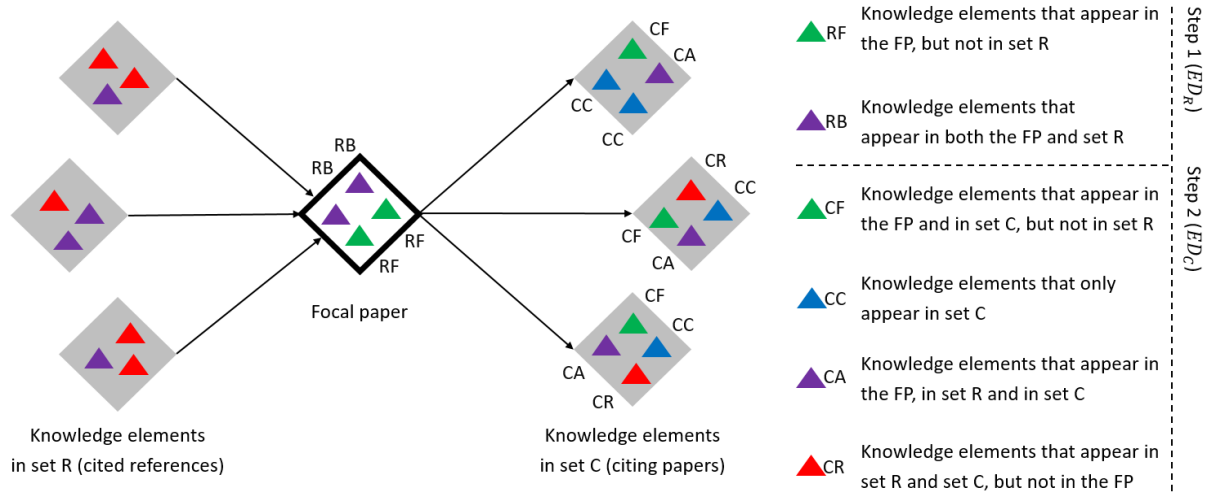

*Figure 6: Graphical representation of an ED network. The illustration is based on S. Wang et al. (2023).*

Instead, it differentiates between six different types of occurrences of knowledge elements within a citation network (Figure 6).

S. Wang et al. (2023) tested two different ways to operationalize knowledge elements. The first approach treats every MeSH term as a knowledge element. This means that the resulting index, referred to as ED$_{(ent)}$, looks for FPs with unique MeSH terms (compared to their cited references). In contrast, the second approach is based on MeSH co-occurrences. Therefore, ED$_{(rel)}$ searches for unique combinations of MeSH terms. Out of all variants of the DI$_1$ (explained so far), ED$_{(rel)}$ shares the most similarities with key-word-based novelty indices.

$$ED_R = \frac{n_{RF} - n_{RB}}{n_{RF} + n_{RB}}; \quad ED_C = \frac{1}{N}\sum_{c=1}^{n}\frac{n_{CF} + n_{CC} - n_{CA} - n_{CR}}{n_{CF} + n_{CC} + n_{CA} + n_{CR}}$$

$$ED = \alpha \times ED_R + (1 - \alpha) \times ED_C$$

The calculation of ED takes three steps. In the first step, ED considers the relationship of the FP to its cited references. ED$_R$ "quantifies the knowledge change directly caused by FP compared to existing research stream" (S. Wang et al., 2023, p. 154) by subtracting the proportion of knowledge elements found in both the FP and its cited references ($RB$) from the proportion of knowledge elements only found in the FP ($RF$).

This procedure is followed up by a second step that groups the knowledge elements contained in the FP's citing papers into one of four categories: "(1) knowledge elements derived exclusively from FP [here: CF]; (2) knowledge elements derived from both FP and its predecessors [here: CA]; 3) knowledge elements only derived from FP's predecessors [here: CR]; and (4) knowledge elements that only appear in the citing publication itself [here: CC]" (S. Wang et al., 2023, pp. 154-155). Like in step one, the number of knowledge elements that originate from the FP's cited references is subtracted from the number of the elements that indicate new and original ideas introduced by either the FP or its citing papers. In the third and last step, the two equations from step one and two are combined using a parameter $\alpha$ that defaults at 0.5 and can be used to give more weight to one part of the equation, if so desired. In fact, S. Wang et al. (2023) recommend using $\alpha < 0.5$ because their results suggest that $ED_C$ contributes more to correct identification of breakthrough papers than $ED_R$. Since the calculation of ED involves multiple steps, the six groups of knowledge elements are not mutually exclusive. For example, one and the same knowledge element may be part of $RF$ in step one and $CF$ in step two.

Following the example of Funk and Owen-Smith (2017), S. Wang et al. (2023) also suggested a second version of the ED that includes a weighting parameter $m_t$. The parameter $m_t$ measures the extent to which the FP's knowledge elements are used by future research and is calculated as follows:

$$m_t = \frac{N_S - N_{\min\_y}}{N_{\max\_y} - N_{\min\_y}} \quad mED = m_t \times ED$$

In this calculation, "$N_S$ is the number of papers that cite FP and share at least one knowledge element with FP at time $t$, $N_{\min\_y}$ is the minimal value of $N_S$ of all papers published in year $y$, and $N_{\max\_y}$ is the maximum value of $N_S$ of all papers published in year $y$" (S. Wang et al., 2023, p. 155). By including $m_t$ in the calculation of ED, the weighted entity-based disruption index (mED) is obtained. The mED measures the time-normalized and knowledge-filtered citation impact of the FP. Both the ED and the mED range from -1 to 1.

## 3.6   Illustration of possible combinations

Since most of the variants of the $DI_1$ do not mutually exclude each other, there are many possible combinations of different variants. For example, merging $DI_{5n}$ with $m_t$ results in $mDI_{5n}$. Because the great diversity of possible combinations makes it impossible to list them all, Table 1 merely serves to hint at the great number of possible index variants. Although not all combinations of indices are equally fruitful, Table 1 gives researchers the option to choose a variant that is tailored to the specific research questions they want to answer. For example, researchers who want to find the most disruptive publications in specific disciplines might find $DI_{5n}$ useful.

*Table 1: Illustration of possible combinations of different variants of the $DI_1$.*

| Indicator | $I = 5$ | $X = 3$ | No $N_R$ | Field-specific | $m_t$ |
|---|---|---|---|---|---|
| $DI_5$ | / | $DI_5^{3\%}$ | $DI_5^{noR}$ | $DI_{5n}$ | $mDI_5$ |
| $DI_{3\%}$ | $DI_5^{3\%}$ | / | $DI_{3\%}^{noR}$ | $DI_{3\%n}$ | $mDI_{3\%}$ |
| $DI_1^{noR}$ | $DI_5^{noR}$ | $DI_{3\%}^{noR}$ | / | $DI_{1n}^{noR}$ | $mDI_1^{noR}$ |
| $DI_{1n}$ | $DI_{5n}$ | $DI_{3\%n}$ | $DI_{1n}^{noR}$ | / | $mDI_{1n}$ |
| $mDI_1$ | $mDI_5$ | $mDI_{3\%}$ | $mDI_1^{noR}$ | $mDI_{1n}$ | / |
| $D$ and $C$ | $D_5; C_5$ | $D_{3\%}; C_{3\%}$ | $D^{noR}; C^{noR}$ | $Dn; Cn$ | $mD; mC$ |
| $DEP$ | $DEP_5$ | $DEP_{3\%}$ | / | $DEP_n$ | $mDEP$ |
| $Orig_{base}$ | $Orig_{base}^5$ | $Orig_{base}^{3\%}$ | / | $Orig_{base}^n$ | $mOrig_{base}$ |
| $ED_{(rel)}$ | $ED_{(rel)}^5$ | $ED_{(rel)}^{3\%}$ | / | $ED_{(rel)}^n$ | $mED_{(rel)}$ |

# 4 Possible disadvantages of using citation data to measure disruption and consolidation

With the exception of the ED, the $DI_1$ and all of its variants rely on citation data to measure disruption and consolidation. For multiple reasons, citation data may not be treated as a perfect representation of the disruptive and consolidating qualities of publications and patents. The citations of patents and scientific publications paint only an incomplete picture of the knowledge and the ideas that circulate through the relevant communities. Not all inventors seek patent protection for their inventions (Funk & Owen-Smith, 2017), and not all publications are properly indexed by bibliometric databases. In the science system, the gap between the total amount of publications and the amount of publications indexed by bibliometric databases is much larger in the social sciences and the humanities than in the natural and life sciences (Bornmann, 2020; Moed, 2005). In summary, this means that there is the danger of selection bias when using citation data to measure disruption and consolidation.

The $DI_1$ and its variants are further limited by the fact that actual citation behaviour is not always in line with the normative citation theory (Merton, 1988), which states that citations represent cognitive influences and are used to give credit to previous research or to prior arts. In reality, however, the inclusion or omission of citations of patents may be a strategic process and some companies may have incentives not to properly cite all prior arts (Alcácer et al., 2009). Similarly, the cited references of a scientific publication often do not represent all of the sources of inspiration that went into a paper (Tahamtan & Bornmann, 2018b). Since citations are a "complex, multidimensional and not a unidimensional phenomenon" (Bornmann & Daniel, 2008, p. 69), any application of the $DI_1$ and its variants will be limited by noisy data.

In addition to the general limitations of citation data, the following subsections provide a summary of studies that examine possible (data-induced) biases that might affect the $DI_1$ and its variants. An unbiased index should only be affected by parameters that relate to the theoretical construct that the index is supposed to measure. In case of the $DI_1$ and its variants, this means that disruption scores should only reflect the disruptive and consolidating qualities of publications (and nothing else). If, on the other hand, parameters that are unrelated to disruption and consolidation affect disruption scores, then it may be concluded that the $DI_1$ and its variants suffer from biases. Each of the following subsections represent a different kind of bias that was investigated in the literature: inconsistency, time-dependency, biases related to reference lists, and coverage-induced biases.

## 4.1 $N_R$ as a source of inconsistency

Consistent disruption indices should have the following feature: Disruptive qualities of an FP should always lead to higher disruption scores and consolidating qualities should always lead to lower disruption scores. With only a few calculations, Wu and Wu (2019) managed to prove that the $DI_1$ as well as many of its variants are not consistent. The inconsistency is caused by the term $N_R$. $N_R$ represents consolidating qualities and is therefore supposed to weaken the disruption score of papers. This works as intended, as long as the numerator $(N_F - N_B)$ is positive. In case, however, that $(N_F < N_B)$ an issue arises: $N_R$ actually strengthens the disruptiveness of papers with negative disruption scores. This problem is illustrated in Table 2. As Table 2 shows, the performance of FP C and FP D is identical with the exception of the $N_R$ values. FP D should be assigned a lower disruption score than FP C, since a high $N_R$ is supposed to indicate consolidation. The results in Table 2 show, however, that $N_R$ artificially

| FP | $N_F$ | $N_B$ | $N_R$ | $DI_1$ |
|----|-------|-------|-------|--------|
| A  | 90    | 10    | 0     | 0.80   |
| B  | 90    | 10    | 100   | 0.40   |
| C  | 10    | 90    | 0     | -0.80  |
| D  | 10    | 90    | 100   | -0.40  |

inflates FP D's disruption score because it strengthens the denominator. Thus, FP D is falsely rewarded with a higher $DI_1$ score than FP C. The issue is caused by the fact that $N_R$ is only part of the denominator and thus has no influence on whether the disruption score is positive or negative: $DI_1 < 0$ if $N_F < N_B$. The same issue also affects $mDI_1$, $DI_l$, $DI_{X\%}$, and $DI_n$. Variants that are positive by definition as well as variants that do not contain $N_R$ do not suffer from the inconsistency. ED is also not affected because every term in the denominator is even part of the nominator: $ED_R < 0$ if $N_{RF} < N_{RB}$; $ED_C < 0$ if $(N_{CF} + N_{CC}) < (N_{CA} + N_{CR})$.

Another consequence of the inconsistency is that $N_R$ pushes the disruption scores of all papers closer to zero, regardless of whether they are on the consolidating or the disruptive part of the scale. Since $N_R$ tends to be quite large in many cases, $DI_1$ assigns values of close to zero to many papers (Bornmann & Tekles, 2021; Leydesdorff et al., 2021). One could argue that this goes against the original intention of Funk and Owen-Smith (2017) to create a nuanced metric because it "raises the question of whether different nuances of disruptions can be adequately captured by $DI_1$, or if the term [here: $N_R$] is too dominant for this purpose" (Bornmann et al., 2020a, p. 1245).

## 4.2 Time-dependency of disruption scores

Since the citation network of an FP keeps changing as long as the FP keeps receiving additional citations, disruption scores may vary greatly depending on the time of measurement. Using four example papers, Bornmann and Tekles (2019a) investigated the variation of $DI_1$ scores over time (Figure 7). While the disruption score of Randall and Sundrum (1999) stabilized rather quickly, it took Davis et al. (1995) five years to arrive at a stable disruption value. The development of the $DI_1$ scores of Oregan and Gratzel (1991) and Iijima (1991) is also insightful, because even after 15 years it seems they still had not fully stabilized. Note that the time-sensitivity affects all citation based variants of the $DI_1$ and not just $DI_1$. Based on their observations, Bornmann and Tekles (2019a) propose a citation window of at least three years, as it is recommended in bibliometrics.

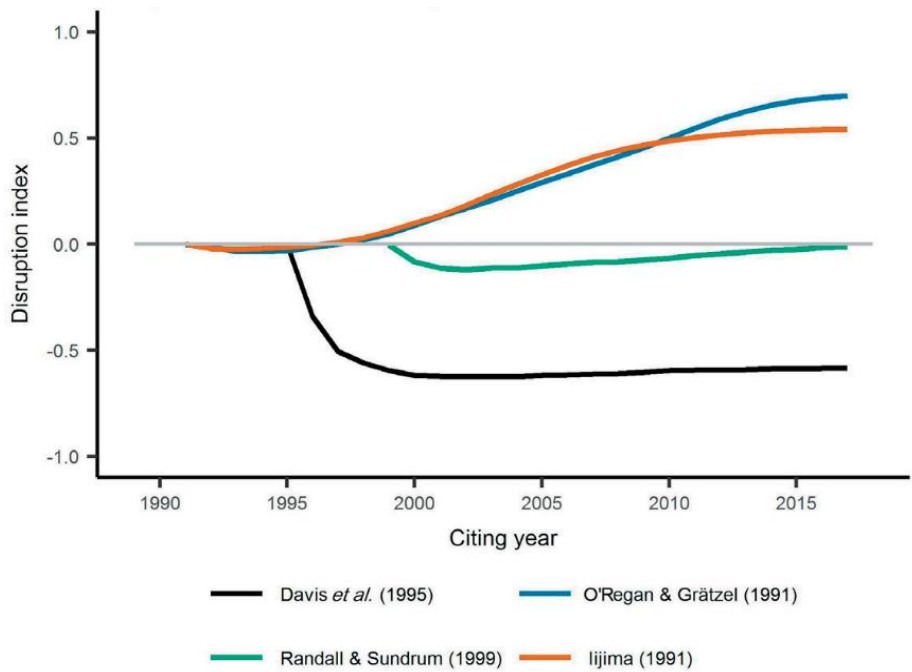

*Figure 7: Variation of four example papers over time (Bornmann & Tekles, 2019a).*

## 4.3   Possible biases caused by the number and the citation impact of cited references

In addition to time-dependency, the $DI_1$ and its variants may also be biased by the total number and the citation impact of the FP's cited references. The more references an FP contains and the more citations these references have received in total, the more likely it is that the citing papers cite at least one of the FP's cited references. Liu et al. (2023) demonstrated the effect that the removal or addition of an important cited reference may have on an FP's $DI_1$ score in a case study on the scientometric literature on so called Sleeping Beauties (SB). SBs are publications that remain mostly unnoticed for a long time and then suddenly attract a lot of attention (van Raan, 2004). For example, the removal of the highly cited paper by van Raan (2004) from the reference list of Ke et al. (2015) significantly changes the $DI_1$ score of Ke et al. (2015). In a dataset consisting of 165 papers from the SB literature, the $DI_1$ score of Ke et al. (2015) increases from -0.541 to -0.176 after the removal of van Raan (2004). The removal of cited publications with less citation impact than van Raan (2004) has little to no effect on the $DI_1$ scores of Ke et al. (2015). Liu et al. (2023) obtained similar findings in a second case study on TD-IDF literature. Because citing a highly cited reference may result in lower $DI_1$ scores, Liu et al. (2023) conclude that according to the $DI_1$ "it is difficult for a focal paper to disrupt highly cited predecessors".

Ruan et al. (2021) provide a detailed examination of the way the number of cited references (ref_num) influences the disruption scores of FPs from five different disciplines in the WoS (engineering, management, mathematics, neurosciences, and plant sciences). The WoS dataset was restricted to articles published between 1954 and 2011 which had received at least ten citations within five years after publication. In addition, Ruan et al. (2021) also examined the disruption scores of publications in the Chinese Social Sciences Citation Index (CSSCI) that were published between 2000 and 2013 and had collected at least five citations within five years after publication. A stepwise regression model shows that ref_num is negatively associated with disruption scores for papers with less than ten cited references (and less than five references in the CSSCI). This negative correlation exists in all of the five sampled WoS disciplines, and it is especially pronounced in the fields of engineering and management as well as in the CSSCI. The association between ref_num and disruption scores stays negative when calculated for papers with more than five references in the CSSCI, but it turns into a positive association for papers with more than ten cited references in all five WoS disciplines.

The size of the positive association varies by discipline and is negligible in the fields of management and engineering. These results are consistent with the findings of Sheng et al. (2023) who examined the correlation ref_num and disruption scores for publications with at least ten references and ten citations in the PubMed Knowledge Graph dataset constructed by Xu et al. (2020). For papers published between 2001 and 2010 and a citation window of five years, an ordinary least squares regression revealed that ref_num is positively associated with disruption scores. A robustness check shows that this positive association is even stronger for papers published between 1961 and 2000, but it turns into a weak negative association when calculated for papers published between 2001 and 2010 using an extended citation window of ten years. Overall, the empirical evidence indicates that the correlation between ref_num and disruption scores is non-linear and affected by discipline, publication age, and citation window.

## 4.4 Possible biases related to insufficient coverage of publications within bibliometric databases

Ruan et al. (2021) found evidence pointing towards an inflation of disruption scores in the discipline of management and in the CSSCI caused by the limited coverage of non-journal and non-English publications. For example "the percentages of the papers with [here: $DI_1$] = 1 are higher in the social sciences, i.e., 3.1% and 27.5% in Management and CSSCI, respectively, whereas the percentages are less than 2.0% in the other four fields" (Ruan et al., 2021). Ruan et al. (2021) illustrate why this result is an artefact using two examples: In the social sciences and humanities, non-journal publications like books are more common than in many other fields (Moed, 2005). Because non-journal publications are insufficiently covered in the WoS, they are not considered in the calculation of disruption scores. So in the extreme case of an FP that only cites books, the FP has $N_F = N_B = 0$ and will thus be assigned a disruption score of 1. Analogously, a publication in the CSSCI that only cites English literature receives a disruption score of 1, because the CSSCI does not cover citations of English publications (as a rule).

Providing a similar line of argument, Liang et al. (2022) point out that the lack of coverage within bibliometric databases affects older publications more severely than more recently published literature. Using bibliometric data from the WoS, they report that many publications from early publication years had a $DI_1$ value that equalled 1 each year after publication, "such as the paper (WoS_ID = WOS:000200263700094) that was published in the year 1937. It was cited two times in the year 1939 and one time in the year 1979 but has no references. So, the results of [here: $DI_1$] for this paper, from the third year to 2016, are all equal to 1. Therefore, it will be better to restrict the analysis to more recent publications" (Liang et al., 2022, p. 5728). Note that the issue of publications wrongfully receiving high disruption scores because of a lack of (indexed) references also affects certain document types that usually contain no or only a few references (e.g. editorials, letters, book reviews, meeting abstracts, etc.). In summary, document-type-, language-, field- and publication-age-dependent lack of coverage may artificially boost disruption values.

# 5 Convergent and predictive validity of $DI_1$ and its variants

As Section 3 illustrates, on the one hand, $DI_1$ and its variants seem to be attractive and versatile tools to measure the disruptiveness of research in empirical studies based on large publication sets. In their influential *Nature* study, Park et al. (2023) ask for the disruptiveness of research in all disciplines over a very long period. Such research questions demand a comprehensive literature database including citation data such as the Web of Science (WoS, Clarivate) and a suitable metric using the data such as the $DI_1$. On the other hand, a time consuming, but detailed examination of publications by experts may lead to more valid results (with respect to assessing disruptiveness), because human judgment may be able to detect the disruptiveness of research in ways that escape a citation based measurement. Therefore, there may be a trade-off between feasibility and validity. A detailed validity examination of the $DI_1$ and its variants is necessary to determine whether this trade-off is worth it.

Several studies have been published that tested the validity of the $DI_1$ and its variants in various ways (Chen et al., 2021; Funk & Owen-Smith, 2017; Li & Chen, 2022; Wu et al., 2019). In order to present the information in a systematic and approachable way, this review focuses on studies that tested the *convergent* validity and/or the *predictive* validity of the $DI_1$ and its variants. Convergent validity addresses the question of whether a metric is positively associated with the construct it is supposed to measure. The convergent validity of a metric may be assessed in two ways: A) Checking how well the results of the metric in question correspond with the results of other metrics that measure the same (or similar concepts) (Forthmann & Runco, 2020). B) Checking how well the results of the metric in question correspond with expert evaluations of the same (or similar) concepts (Kreiman & Maunsell, 2011). "The criteria for convergent validity would not be satisfied in a bibliometric experiment that found little or no correlation between, say, peer review grades and citation measures" (Rowlands, 2018). In the case of the $DI_1$ and its variants, the basic idea is to use lists of landmark publications (e.g., publications leading to the Nobel prize, NP) that were compiled by groups of experts and to compare the disruption scores of landmark and non-landmark publications. If disruption scores do measure what they are supposed to measure, they should identify the landmark papers picked by experts by assigning significantly higher scores to landmark publications than to non-landmark publications.

Tests of predictive validity involve the usage of historical data in order to assess how well a bibliometric index is able to predict future outcomes of the concept of interest (Kreiman & Maunsell, 2011). Currently, there is only one study[8], namely Shibayama and Wang (2020), that investigated the predictive validity of an index measuring disruption (i.e. the Shibayama-Wang originality). The methods and results of validation studies are listed and explained in the following subsections. To make the results of the validation studies more easily comparable, the subsections are based on the type of data used to validate disruption indices. Particular focus is placed on the comparative performance of different disruption index variants.

## 5.1 Nobel Prize-winning publications

Wu et al. (2019) were the first to use publication data on NP-winning papers to assess the validity of the $DI_1$. This approach rests on the assumption that papers worthy of a NP are on average more disruptive than other papers. In consequence, the $DI_1$ and its variants are expected to assign higher average disruption values to NPs than to non-NPs. Wu et al. (2019) compared the $DI_1$ scores of 877 NPs published between 1902 and 2009 to a control group of 3,372,570 WoS papers from the same journals and the same years. In accordance with their expectation they found that the average disruption value of NPs is 0.10, placing NPs among the top 2% most disruptive papers out of all WoS publications from the same period. Therefore, Wu et al. (2019) conclude that the $DI_1$ shows favourable (the expected) results.

Following the example of Wu et al. (2019), four later studies also adopted the NP-based approach to evaluate the convergent validity of the $DI_1$ and its variants. All four studies used a (modified version of a) dataset that was provided by Li et al. (2019) and contains "publication records for almost all Nobel laureates in physics, chemistry, and physiology or medicine from 1900 to 2016" (Li et al., 2019). The results of the four studies based on the Nobel laureate dataset are displayed in Table 3. The detailed findings of S. Wang et al. (2023) are displayed separately in Table 4 because it is the only study that tested the convergent validity of multiple disruption indices.

Wei et al. (2020) compared the average $DI_1$ scores of 557 NPs with a control group of 557 randomly sampled non-NPs from the issue of the same journal using t-tests. For the time span from 1900 to 2016, they did not find statistically significant differences between the average disruption values of NPs and non-NPs. This finding stands in contrast to the results of Wei et

---

*Table 3: Results from four validation studies of the DI$_1$ based on NPs.*

| Study | Range of publications years | Statistical procedure | Results for DI$_1$ | Results match expectations |
|---|---|---|---|---|
| *(Wu et al., 2019)* | 1902-2009 | Descriptive examination | Mean NP = 0.10 <br> NPs ranked among the top 2% most disruptive papers in the entire WoS | ✓ |
| *(Wei et al., 2023)* | 1964-2000 | Linear regression | DI$_1$ (coefficient) = 0.071 <br> p<0.001 <br> Citations (coefficient) = 880.946 <br> p<0.001 | ✓ |
| *(Wei et al., 2020)* | 1900-2016 | t-test | Mean (NP) = 0.236 <br> Mean (non-NP) = 0.239 <br> p=0.898 | ✗ |
| *(S. Wang et al., 2023)* | 1991-2014 | Logistic regression | DI$_1$ (coefficient) = -6.777 <br> p=0.125 | ✗ |

al. (2023), who report that the average DI$_1$ values of NPs exceed those of comparable non-NPs by 0.070. The differences in the results may come from differences in the statistics used. Wei et al. (2023) applied a multivariate linear regression, which included some control variables as well as citation counts as an additional dependent variable, to a sample of 164 NPs and 9,034 non-NPs published between 1964 and 2000. Instead of a random sample, Wei et al. (2023) chose to include all non-NPs that were published in the issue of the same journal and in the same year as the NPs. They found that ceteris paribus the average DI$_1$ score of NPs is 0.071 points higher than the average DI$_1$ score of non-NPs. Furthermore, NPs receive significantly more citations (around 881 on average) than non-NPs. As a result, most NPs fall under the "revolutionary science" category in the framework of Wei et al. (2023).

S. Wang et al. (2023) used a similar, but modified dataset. In their own words, their dataset consists of "key publications" (S. Wang et al., 2023, p. 157) from Nobel Prize laureates in the fields of chemistry, medicine, and physics as well as winners of the Lasker Award. The Lasker

*Table 4: Results from logistic regressions based on S. Wang et al. (2023). Pseudo $R^2$ measures the improvement of the prediction compared to the null model only containing the regression constant (and none of the independent variables). The small Pseudo $R^2$ values indicate that disruption values contribute little to the prize status of publications.*

| Index | Coefficient | p-value | Pseudo $R^2$ |
|---|---|---|---|
| $mED_{(rel)}$ | -10.522 | 0.484 | 0.001 |
| $mED_{(ent)}$ | -12.682 | 0.410 | 0.001 |
| $mDI_1$ | -0.024 | 0.042 | 0.007 |
| $DI_5$ | -30.012 | 0.000 | 0.038 |
| $DI_1$ | -6.777 | 0.125 | 0.004 |

Award is given out annually since 1945 and the award program was created "to shine a spotlight on fundamental biological discoveries and clinical advances that improve human health"[9]. The dataset contained 268 prize papers published between 1991 and 2014. For every prize paper, five non prize-winning papers were randomly sampled from PubMed. The non-prize-winning publications had to meet the following criteria: "published in the same year […] with approximately equal citation counts (±5) and the same number of co-authors" (S. Wang et al., 2023, p. 156). A logistic regression was conducted for five indices ($mED_{(rel)}$, $mED_{(ent)}$, $mDI_1$, $DI_5$, $DI_1$) with the prize status as the binary dependent variable (either prize winning or non-prize-winning paper).

The results of the regression analyses are shown in Table 4. Only the coefficients of $DI_5$ and $mDI_1$ achieve statistical significance. Against expectation, all indices have negative coefficients: the higher the disruption value of a publication, the lower the likelihood of it being a prize-winning paper. In other words, prize-winning papers appear to be more consolidating on average than non-prize-winning PubMed publications with comparable bibliometric features.

A hint on how the conflicting findings of the above mentioned studies can be interpreted can be found in the study of Liang et al. (2022). They investigated how the $DI_1$ values of 646 NPs changed over the course of time. The NPs were compared to 653 benchmark papers (BPs) from the same journal and publication year. Since Liang et al. (2022) used the entirety of the dataset provided by Li et al. (2019), their sample contained papers published between 1900 and 2016. Three important conclusions can be drawn from Liang et al. (2022): First, the $DI_1$ values of both NPs and non-NPs vary across disciplines. For example, NPs in the field of medicine tend to be more consolidating than NPs in the fields of physics, and they require more time to achieve positive disruption values. Second, the $DI_1$ values of NPs are very time-sensitive because they tend to keep on accruing citations long after publication. Third, NPs published after 1980 require more time to reach positive disruption values than NPs published before 1980 (Figure 8). The empirical evidence by Liang et al. (2022) and Ruan et al. (2021) (presented in Section 4.4) calls into question the reliability of the results obtained by Wu et al. (2019) and Wei et al. (2020) because the studies included publications that date back to very early publication years. The results of S. Wang et al. (2023) and Wei et al. (2023) are probably more reliable since they are based on publications from more recent publication years. In summary, the results from the research on NPs are inconclusive insofar as it remains unclear whether the $DI_1$ and its variants are able to identify NP-winning research. The current state of research on disruption indices highlights that disruption values of highly cited publications (like NPs) are strongly dependent on sample composition, especially with regard to discipline, publication age, and citation window. Different choices regarding these important factors may lead to "different, even controversial, results" (Liang et al., 2022, p. 5728).

---

[9] https://laskerfoundation.org/awards/about-the-awards/

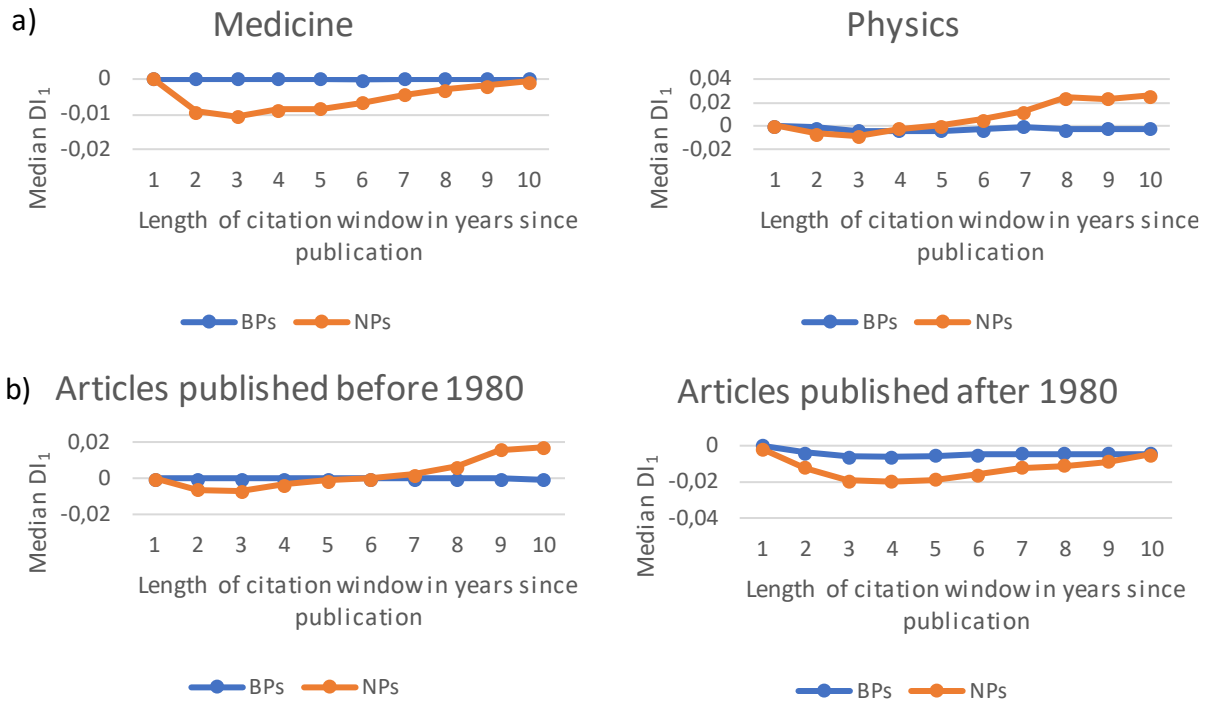

*Figure 8: Median $DI_1$ values for a) articles in the fields of medicine and physics b) articles published before and after 1980 depending on the citation window. The illustration is based on Liang et al. (2022).*

## 5.2 Faculty Opinions

*Faculty Opinions* aims to provide curated selections of relevant and high quality research within the disciplines of biology and medicine. The company is responsible for the *Faculty Opinions* database (formerly known as *F1000Prime*), which is based on a post publication peer review process. Peer-nominated Faculty Members (FMs) rate and rank papers according to their quality and their importance using a three star system (1 star = "good", 2 stars = "very good", 3 stars = "exceptional"). We found three studies that take advantage of *Faculty Opinions* in order to assess the validity of the $DI_1$ and its variants. These studies are Bornmann et al. (2020a), S. Wang et al. (2023) and Wei et al. (2023). Each study took a slightly different approach to operationalizing expert judgements of disruptiveness (and similar concepts).

Bornmann et al. (2020a) used tags, which FMs may choose to assign to papers in addition to ratings. The purpose of the tags is to provide an "'at a glance' guide for the reason(s) the article is being recommended"[10]. Examples of such tags are displayed in Table 5. Bornmann et al. (2020a) expected the index scores to show positive correlations with tags that represent aspects of novelty and show negative correlations with tags that indicate consolidating research. "As disruptive research should include elements of novelty, we expect that the tags are positively related to the disruption indicator scores. For instance, we assume that a paper receiving many 'new finding' tags from FMs will have a higher disruption index score than a paper receiving only a few tags (or none at all)" (Bornmann et al., 2020a, p. 1247). The study was based on a dataset of 157,020 papers published between 2000 and 2016. Only papers with at least ten cited references and at least ten citations were included and a citation window of at least three years was chosen. In total, the $DI_1$ and four variants were tested: $DI_1$, $DI_5$, $DI_1^{noR}$, $DI_5^{noR}$, and DEP. Table 5 shows the results of 45 Poisson regressions. The important aspect here is the sign of the coefficients and not their size. Although the difference in the performance between the index variants is not very large, $DI_5$ produces the best results because six of its coefficients point in the expected direction. DEP shows the weakest performance out of all five indices with only four out of eight coefficients pointing in the expected direction.

*Table 5: Results of 45 Poisson regressions with tags as dependent variables and the $DI_1$ and four variants as independent variables. The regressions are adjusted for exposure time, i.e. publication year: How long have the papers been at risk of being exposed? Results in line with expectations are highlighted in bold. The table is based on Bornmann et al. (2020a).*

| Tag | $DI_1$ | $DI_5$ | $DI_1^{noR}$ | $DI_5^{noR}$ | DEP |
|---|---|---|---|---|---|
| (Expecting positive signs) | | | | | |
| *Hypothesis* | **4.32** | **3.01** | **4.66** | **2.75** | 0.25 |
| *New finding* | -2.71 | -0.62 | -2.13 | -2.13 | 1.92 |
| *Novel drug target* | **6.89** | **6.74** | **14.85** | **15.19** | **-7.91** |
| *Technical advance* | **6.72** | **20.65** | **18.34** | **19.80** | **-18.11** |
| (Expecting negative signs) | | | | | |
| *Confirmation* | **-5.08** | **-0.41** | 0.08 | 1.45 | -1.88 |
| *Good for teaching* | 12.24 | 0.37 | 8.82 | 3.73 | **6.41** |
| *Negative/Null results* | 3.45 | **-11.46** | **-13.35** | **-5.20** | -2.10 |
| *Refutation* | **-9.28** | **-9.59** | **-14.83** | **-10.37** | 8.06 |
| (No expectations) | | | | | |
| *Controversial* | -0.33 | 4.42 | 1.46 | 1.46 | -3.62 |

---

[10] https://facultyopinions.com/faq

*Table 6: Rotated factor loadings from a factor analysis using logarithmized variables [log(y +1)]. The results matching expectations are highlighted in bold. ReSc.Sum is the sum of reviewer scores and ReSc.avg is the average reviewer score. The table is based on (Bornmann et al., 2020a).*

| Variable | Factor 1 (disruption indicator) | Factor 2 (citation impact) | Factor 3 (reviewer scores) |
|---|---|---|---|
| $DI_1$ | 0.24 | **-0.69** | 0.05 |
| $DI_5$ | **0.90** | -0.07 | 0.00 |
| $DI_1^{noR}$ | **0.90** | -0.10 | 0.02 |
| $DI_5^{noR}$ | **0.97** | -0.03 | 0.01 |
| DEP | **-0.91** | -0.01 | 0.01 |
| Citations | 0.05 | **0.91** | 0.04 |
| Citation impact percentiles | 0.04 | **0.84** | 0.12 |
| ReSc.sum | 0.00 | 0.05 | **1.00** |
| ReSc.avg | 0.00 | 0.05 | **1.00** |

Bornmann et al. (2020a) also investigated the relationship between indices measuring disruptiveness, citation impact, and reviewer scores by calculating a factor analysis (FA). FA is an explorative method designed to identify latent dimensions in a given dataset (Baldwin, 2019; Gaskin & Happell, 2014). Bornmann et al. (2020a) expected to find three dimensions in the data. The FA does indeed identify three dimensions (Table 6), but it also reveals two surprising results: "[C]ontrary to what was expected, $DI_1$ loads negatively on the citation dimension revealing that (a) high $DI_1$ scores are related to low citation impact scores […] and (b) all other indicators measuring disruption are independent of $DI_1$" (Bornmann et al., 2020a, p. 1252). In other words, $DI_5$, $DI_1^{noR}$, $DI_5^{noR}$, and DEP load strongly on the same dimension, implying that at least one of them is an improvement compared to the $DI_1$. The FA also shows that disruption scores do not correlate with reviewers' ratings, suggesting that reviewers do not tend to assign higher (or lower) star ratings to disruptive publications than to consolidating publications.

*Table 7: Results from logistic regression analysis based on S. Wang et al. (2023). Results that match expectations are highlighted in bold.*

| Index | Coefficient | Pseudo $R^2$ |
|-------|-------------|--------------|
| $mED_{(rel)}$ | **93.749 (0.000)** | 0.045 |
| $mED_{(ent)}$ | **35.731 (0.000)** | 0.005 |
| $mDI_1$ | -0.022 (0.001) | 0.005 |
| $DI_5$ | -4.999 (0.000) | 0.002 |
| $DI_1$ | -5.650 (0.000) | 0.003 |

The lack of a correlation between reviewers' ratings and disruption scores probably affects the results of S. Wang et al. (2023), who assessed the validity of $DI_1$, $mDI_1$, $DI_5$, $mED_{(ent)}$, and $mED_{(rel)}$ using a combination of tags and reviewers' ratings. They categorized papers that earned a reviewer score of at least two stars and received the tags "Hypothesis", "New finding", "Novel drug target", "Technical advance", and "Changes in clinical practice" as breakthrough papers. S. Wang et al. (2023) collected 2,002 breakthrough papers that were published between 1991 and 2002. They constructed a dependent variable that is 1 if the paper is a breakthrough paper and 0 if not. A logistic regression analysis was performed with the $DI_1$ and its variants as independent variables. The authors tested whether the indices are able to differentiate between breakthrough papers and randomly sampled non-breakthrough papers from the PubMed database with the same publication year and similar citation counts. The results of the logistic regression analyses are displayed in Table 7. In contrast to the results obtained by Bornmann et al. (2020a), $DI_5$ shows poor performance and $DI_1$ performs even worse. On average, $DI_5$ and $DI_1$ assigned lower disruption values to breakthrough papers than to non-breakthrough papers. Out of all variants, only $mED_{(rel)}$ and $mED_{(ent)}$ produced coefficients that pointed in the expected direction. The results also show that the index based on MeSH co-occurrences ($mED_{(rel)}$) is significantly more effective than the index based on unique mesh terms ($mED_{(ent)}$). In line with this observation, the Pseudo $R^2$ values also indicate that $mED_{(rel)}$ performs best out of all five indices at identifying breakthrough papers.

*Table 8: Examples of reviewers' comments based on Wei et al. (2023). Words that point to revolutionary science are underlined. Although the second comment does not contain any of the words listed by Wei et al. (2023), the paper was still coded as "revolutionary science".*

| | Review Comment |
|---|---|
| 1 | I have found that it is such an outstanding article overall. I find the work <u>innovative</u> and recommend indexing. |
| 2 | This paper provides an important advance in the study of spatial proteomics. |
| 3 | EGSEA is a new gene set analysis tool that <u>combines</u> results from multiple individual tools in R as to yield better results. The authors have published the EGSEA methodology previously. This paper focuses on the practical analysis workflow based on EGSEA with specific examples. As EGSEA is a compound and complicated analysis procedure, this work serves as valuable guidance for the users to make full use of this tool. |

Unlike Bornmann et al. (2020a) and S. Wang et al. (2023), Wei et al. (2023) focused on reviewers' comments instead of tags. They decided to recognize a paper as "revolutionary science" if its review comments included the words "innovative", "revolutionize", "revolutionary", "novel", "novelty", "creativity", "creative", "innovation", "original", "initial", "originality", "radical", "breakthrough", "new", "bridge", "combine", "first ones", "contribute to", "thought-provoking" or "provocative". They provided examples of reviews that indicate revolutionary, i.e. disruptive, science (Table 8). Wei et al. (2023) collected 70 revolutionary papers and compiled a control group of 1,405 papers from the same journals.

Based on the reviewers' comments, Wei et al. (2023) created a binary variable that assumes 1 if a paper is considered revolutionary science and assumes 0 otherwise. This binary variable was used as the independent variable in a multivariate linear regression. The dependent variables were citation counts and $DI_1$ values; the control variables were number of authors and number of cited references. The analysis reveals that both the average citation counts and the average $DI_1$ values are higher for revolutionary papers than for papers in the control group (25 citations and 0.016 value points, respectively). Both coefficients are statistically significant and point in the expected direction, but the coefficient for disruption scores is very small. Wei et al. (2023) conclude that the combination of citation counts and $DI_1$ is able to correctly identify disruptive science. In summary, the $DI_1$ and its variants perform significantly less favourable in the study of S. Wang et al. (2023) than in the studies of Bornmann et al. (2020a) and Wei et al. (2023). A short overview of the results is provided in Table 9. The weak correlation between disruptions cores and reviewer scores observed in the FA by Bornmann et al. (2020a) may have caused $mDI_1$, $DI_1$ and $DI_5$ to perform poorly in the calculations of S. Wang et al. (2023). While the $DI_1$ and its variants are not able to identify publications with high reviewer scores, $DI_5$ and DEP in particular seem capable of identifying novel research.

*Table 9: Short overview of the results from studies based on Faculty Opinions.*

| Study | Operationalization | Statistical procedure | Favorable performance | Least favorable performance |
|---|---|---|---|---|
| *Bornmann et al. (2020a)* | Tags | Poisson regression | $DI_5$ | DEP |
| *Wei et al. (2023)* | Comments | Multivariate regression | $DI_1$ (+ citation counts) | / |
| *S. Wang et al. (2023)* | Tags and scores | Logistic regression | $mED_{(rel)}$, $mED_{(ent)}$ | $DI_5$, $DI_1$ |

## 5.3   Milestone and breakthrough papers

In 2008, *Physical Review Letters* (PRL) compiled a list of milestone publications to celebrate the journal's 50[th] anniversary. The list includes publications from 1958 to 2001 and is available online.[11] According to the information provided by the publisher on this website, the collection "contains Letters that have made long-lived contributions to physics, either by announcing significant discoveries, or by initiating new areas of research. A number of these articles report on work that was later recognized with a Nobel Prize for one or more of the authors". The milestones papers have been carefully selected to represent the various subdisciplines of the field of physics. A similar list of milestone papers was published in 2015 by *Physical Reviews E* (PRE) in celebration of its 50,000[th] publication.[12] This collection includes papers published between 1993 and 2004 and with the objective to identify significant scientific contributions in different fields of physics.

---

[11] https://journals.aps.org/prl/50years/milestones

As of now, there are three studies that have used the PLR and PRE collections to assess the convergent validity of the $DI_1$ and its variants. Based on the assumption that milestone assignments are a proxy for disruption, the three studies investigated whether the $DI_1$ and its variants assign higher values to milestone than to non-milestone publications. We will start with two studies that are very similar: Bornmann and Tekles (2021) and Bittmann et al. (2022). Both studies tested the same set of indices ($DI_1$, $DI_5$, $DI_{1n}$, $DI_{5n}$, and DEP) as well as the same set of control variables (number of co-authors, number of countries, number of cited references, and publication year). They also employed coarsened exact matching (CEM) to combat issues caused by unbalanced data. Bittmann et al. (2022) collected a total of 21,153 ordinary papers and 21 milestone papers from the PRE dataset. Bornmann and Tekles (2021) restricted the PLR dataset to 44,812 articles published between 1980 and 2002. Out of these 44,812 articles only 39 have been classified as milestone papers by the journal's editors. In both studies, a small number of milestone papers is compared to a massive number of non-milestone papers resulting in highly unbalanced data.

In cases where observable covariates are unevenly distributed among unbalanced data, matching algorithms may be used to combat biases: "The general idea behind statistical matching is to simulate an experimental design when only observational data are available to make (causal) inferences. In an experiment, usually two groups are compared: treatment and control. The randomized allocation process in the experiment guarantees that both groups are similar, on average, with respect to observed and unobserved characteristics before the treatment is applied. Matching tries to mimic this process by balancing known covariates in both groups. The balancing creates a statistical comparison where treatment and control are similar, at least with respect to measured covariates" (Bittmann et al., 2022, p. 1251). In other

---

*Table 10: Results from CEM based on Bittmann et al. (2022) and Bornmann and Tekles (2021). The standard errors are displayed in brackets. Note that Bittmann et al. (2022) multiplied $DI_1$, $DI_5$, $DI_{1n}$, and $DI_{5n}$ by 100 to avoid small numbers with many decimal places.*

| Variable | ATE | ATE |
| --- | --- | --- |
| | *Bittmann et al. (2022)* | *Bornmann and Tekles (2021)* |
| $DI_1$ | 3.0627 | 0.14959* |
| | (2.9518) | (0.06192) |
| $DI_5$ | 7.9726* | 0.23884*** |
| | (3.2481) | (0.05506) |
| $DI_{1n}$ | -0.0148 | -0.00033 |
| | (0.0119) | (0.00024) |
| $DI_{5n}$ | 0.0209 | 0.00175** |
| | (0.0123) | (0.00053) |
| *DEP (inverse)* | 1.7352*** | 0.51217*** |
| | (0.2127) | (0.12102) |
| *Logarithmized citation counts* | 3.7807*** | 4.03215*** |
| | (0.1289) | (0.18535) |

*p < .05, **p < .01, ***p < .001

words, if there is covariate $z$ that is unequally distributed among treatment and control group, a matching algorithm tries to compare members of the treatment and control group with comparable z-values.

Since both Bornmann and Tekles (2021) and Bittmann et al. (2022) mainly relied on CEM, we forgo a detailed explanation of the different types of matching algorithms and focus on CEM instead. Unlike other matching algorithms, CEM tries to find exact matches and actively discards dissimilar cases from the calculation. This procedure considerably improves the balancing of the data. If perfect matches are difficult to find, coarsening is employed: "For example, a continuous variable with a large number of distinct values is coarsened into a prespecified number of categories, such as quintiles. Matching is then performed based on quintile categories, and the original information is retained. After matching based on the coarsened variables, the final effects are calculated as differences in the outcome variable between group means using the original and unchanged dependent variable" (Bittmann et al., 2022, p. 1255).

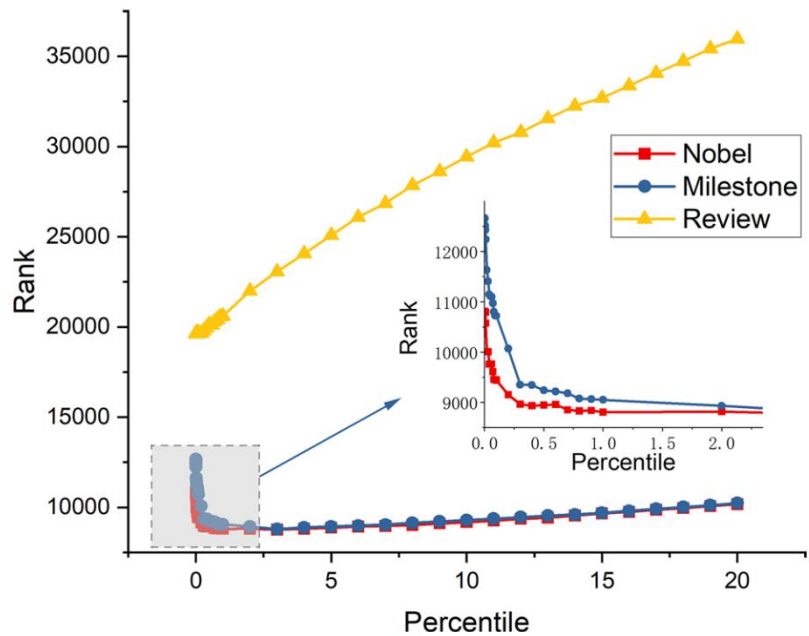

*Figure 9: Illustration of how the choice of X for $DI_{X\%}$ influences the ranks of review articles, milestone papers, and NPs (Deng and Zeng, 2023).*

The results of Bornmann and Tekles (2021) and Bittmann et al. (2022) are displayed side by side in Table 10. The average treatment effect (ATE) is the estimated difference in disruption values between milestone publications and ordinary (control) publications. For example, in Bornmann and Tekles (2021) the ATE is 0.14959 for $DI_1$. This means that the average $DI_1$ value of milestone publications exceeds the average $DI_1$ value of ordinary PRE publications by about 0.15 points. While $DI_5$ and DEP show promising results, $DI_{5n}$ and $DI_{1n}$ fail to differentiate between milestone and control papers. It is also worth noting that milestone papers receive more citations on average than the publications in the control group, which is why logarithmized citation counts can identify milestone papers well. Regarding effect size and effect direction, the results of Bornmann and Tekles (2021) correspond with the findings of Bittmann et al. (2022). While the ATEs of $DI_1$ and $DI_{5n}$ in Table 10 point in the expected direction, they cannot compete with the convergent validity of DEP and $DI_5$. The coefficient for DEP is very large in the results of Bittmann et al. (2022), indicating that DEP is able to detect the disruptive qualities of milestone papers better than other variants of the $DI_1$.

The third study using the PLR dataset was conducted by Deng and Zeng (2023). They collected 87 PLR milestone papers. Out of these publications, 49 are NPs. Furthermore, 1,671 review articles published in *Review of Modern Physics* were selected to serve as representations of consolidating research. Deng and Zeng (2023) compared the disruption values of three groups of publications, i.e., NPs, milestone papers, and review articles, and analyzed ordinary papers in the database of the *American Physiological Society*. After restricting the sample to papers that received at least five citations (to include papers with a minimum impact), the sample consisted of 230,867 publications.

*Table 11: Average percentile ranks of three types of publications depending on the choice of index based on Deng and Zeng (2023). According to Deng and Zeng (2023), t-tests showed that the differences between the three indices are statistically significant for milestone papers and NPs (p<0.01). For review letters, only the difference between DI$_5$ and DI$_1$ reached statistical significance (p < 0.01; p = 0.855 for DI$_1$ and DI$_{3\%}$).*

| Type of publication | DI$_1$ | DI$_5$ | DI$_{3\%}$ |
| --- | --- | --- | --- |
| Milestone | 63.88 | 83.59 | 92.72 |
| Nobel Prize | 69.67 | 86.91 | 92.40 |
| Review Letter | 64.77 | 58.95 | 64.69 |

Unlike Bornmann and Tekles (2021) and Bittmann et al. (2022), Deng and Zeng (2023) chose not to calculate ATEs and instead opted for a different approach. They assumed that review papers are prime examples of consolidating publications, and that milestone papers and NPs are the most disruptive papers in the dataset. In their study, disruption was measured with ranks instead of scores. For example, a paper ranked at 15 has a higher disruption value than a paper ranked at 200. Therefore, a valid index should (1) assign lower ranks (i.e., ranks closer to 0) to milestone papers and NPs than to ordinary papers and (2) assign higher ranks to reviews than to primary research papers. Three indices were tested: DI$_1$, DI$_5$, and DI$_{X\%}$. Figure 9 shows how the percentile choice for DI$_{X\%}$ changes the ranks of the publications. Deng and Zeng (2023) report that a threshold of 3% produces the best results because DI$_{3\%}$ assigns minimum ranks to milestone papers and NPs.

Table 11 provides an overview of the performances of the three indices. It displays percentile ranks instead of absolute ranks. Low percentile ranks indicate consolidation, and high percentile ranks indicate disruptive science. DI$_5$ succeeds in assigning particularly low percentile ranks to review letters, while DI$_{3\%}$ maximizes the percentile ranks of NPs and milestone papers. Therefore, Deng and Zeng (2023) concluded that DI$_5$ as well as DI$_{3\%}$ are considerable improvements compared to DI$_1$. DI$_5$ appears to excel at the identification of consolidating research, whereas DI$_{3\%}$ performs well at identifying disruptive research.

Following the same fundamental idea as the three studies above, but using a different dataset, S. Wang et al. (2023) tested the ability of five indices (mED$_{(rel)}$, mED$_{(ent)}$, mDI$_1$, DI$_5$, DI$_1$) to identify articles that have been declared as breakthrough papers by the *Science* magazine.

*Table 12: Results from logistic regressions based on S. Wang et al. (2023). p-values are displayed in brackets. Results that match expectations are highlighted in bold.*

| Index | Coefficient | Pseudo $R^2$ |
|---|---|---|
| $mED_{(rel)}$ | **62.849 (0.000)** | 0.024 |
| $mED_{(ent)}$ | **54.041 (0.010)** | 0.010 |
| $mDI_1$ | -0.018 (0.087) | 0.005 |
| $DI_5$ | -0.660 (0.804) | 0.000 |
| $DI_1$ | -7.449 (0.047) | 0.007 |

In 1989, *Science* magazine started awarding the title of "molecule of the year" to molecules connected to major scientific developments (Guyer & Koshland, 1989). In 1996, the award was renamed to "breakthrough of the year". It is given to publications that made significant contributions to a field of research and led to a scientific breakthrough (not necessarily connected to molecules). From this annually updated list of breakthrough publications, S. Wang et al. (2023) collected 321 articles published between 1991 and 2014. A logistic regression reveals that only $mED_{(rel)}$ and $mED_{(ent)}$ are able to identify the breakthrough papers among PubMed publications with comparable bibliometric features (Table 12). The performance of $mED_{(rel)}$ is clearly superior compared to any other index.

## 5.4 Self-assessments of researchers

Shibayama and Wang (2020) tested both the convergent and the predictive validity of $Orig_{base}$, $Orig_{weighted\_yc}$, and $Orig_{weighted\_zr}$ by combining survey data with bibliometric data from the WoS. The survey was mailed to 573 randomly selected researchers in the life sciences who had earned their PhD degrees between 1996 and 2011 in Japan. Shibayama and Wang (2020) used a two-dimensional concept of originality and differentiated between theoretical and methodological originality. The respondents were asked to the evaluate the theoretical as well as the methodological originality of their dissertation projects using a three-point scale: 0 (not original), 1 (somewhat original), 2 (original). The survey data was then linked with bibliometric data about the publications that the researchers had published in the year of their graduation or 1-2 years before. In total, 246 responses from the survey and the bibliometric data for 546 publications were collected.

The test of the convergent validity was conducted by calculating correlation coefficients between originality scores and self-assessed originality to check how well the variants of the Shiayama-Wang originality correspond with the self-assessments of researchers. The strength of the correlation was tested under different conditions by varying the length of the citation window and the amount of citing papers considered in the calculation of the Shibayama-Wang originality. The results show that the Shibayama-Wang originality performs significantly better at identifying theoretical innovations than methodological innovations. For theoretical innovations, the correlation coefficients are mostly statistically significant and range from around 0.10 to around 0.17 (depending on the test conditions). The coefficients for methodological innovations, however, are either negative or do not reach statistical significance. The performance of $Orig_{base}$, $Orig_{weighted\_yc}$, and $Orig_{weighted\_zr}$ was rather similar overall and no variant of the Shibayama-Wang originality displayed clearly better convergent validity than the other variants.

*Table 13: Results from logistic regression based on Shibayama and Wang (2020).*

| Indicator | Coefficient | Standard error | p-value |
|-----------|-------------|----------------|---------|
| $Orig_{base}$ | 9.316 | 3.108 | p<0.01 |
| $Orig_{weighted\_yc}$ | 13.167 | 7.192 | p<0.1 |
| $Orig_{weighted\_zr}$ | 35.700 | 15.480 | P<0.05 |

To assess the predictive validity of disruption indices, Shibayama and Wang (2020) investigated whether $Orig_{base}$, $Orig_{weighted\_yc}$, and $Orig_{weighted\_zr}$ are able to predict highly impactful papers. The design of this test was guided by the notion that highly original research is more likely to be highly cited than less original research. Based on citation counts up to 2018 a binary variable was constructed that assumes 1 if a paper is among the top 10% most highly cited publications in the dataset and assumes 0 otherwise. A logistic regression showed that all variants of the Shibayama-Wang originality are able to predict future citation impact reasonably well and similarly well (Table 13).

In summary, $Orig_{base}$, $Orig_{weighted\_yc}$, and $Orig_{weighted\_zr}$ displayed similar levels of convergent and predictive validity. For the most part, the findings of Shibayama and Wang (2020) seem to confirm the validity of the Shibayama-Wang originality. However, there are some limitations. According to Funk and Owen-Smith (2017), disruption indices measure a construct that may not be adequately captured by impact measures. Therefore, it might not be ideal to use impact measures to test the predictive validity of disruption metrics. Furthermore,

Shibayama and Wang (2020) acknowledge that their study is limited to bibliometric data from the life sciences. Since different disciplines exhibit different citing behaviour (and might also have different field-specific standards regarding the assessment of novelty), the results of Shibayama and Wang (2020) might not be applicable outside of the life sciences. All in all, the limitations highlight the need for further research on the matter.

# 6 Discussion

There are two main takeaways from the current research on disruption indices. On the one hand, the $DI_1$ and its variants enable the exploration of intriguing research questions that require vast amounts of bibliometric data. Two popular bibliometric studies published by Wu et al. (2019) and Park et al. (2023) in *Nature* would scarcely be possible without the $DI_1$. On the other hand, it is apparent that the $DI_1$ and its variants have some considerable limitations that they share with many other bibliometric indices. Not only are these indices, as citation-based indices, highly time-sensitive metrics and dependent on many factors that influence citation decisions (e.g., the language of the cited paper or the reputation of the authors) (Tahamtan & Bornmann, 2018a). Disruption scores are also heavily affected by a time-, discipline-, document type-, and language-related lack of coverage in literature databases. Because no database offers perfect coverage of the worldwide literature and because coverage is generally worse for publications from early publication years, there is no way to rule out the possibility that the results of historic studies like Park et al. (2023) may (in part) be distorted (despite the study's robust methodology): Early publications may have artificially inflated disruption scores.

 Not all variants of the $DI_1$ are concerned by the limitations in the same way. Many different variants have been proposed in recent years. For example, the ED and mED do not suffer from the usual limitations of citation data, since they are based on MeSH terms. These deviations from the $DI_1$ and several variants, however, do not only lead to advantages; they also lead to specific limitations. MeSH terms, as used for the ED and mED, are a reliable source of key words to identify the content of research, but their availability is limited to PubMed publications. However, the indices cannot be computed for publications (focal publications as well as cited and citing publications) without standardized key words such as MesH terms in PubMed.

## 6.1 Convergent and predictive validity

A systematic review of the literature reveals that empirical evidence on the predictive validity of disruption indices is still too scarce to arrive at any substantial conclusions. The current literature only provides some illustrative calculations with the D and C indices (Li & Chen, 2022) and more detailed evidence on the predictive validity of the Shibayama-Wang orginality (Shibayama & Wang, 2020). Compared to predictive validity, there is a richer body of literature on the convergent validity of disruption indices, but the results are not entirely conclusive. There are only two consistent findings across all studies: 1) Citation impact measures are strongly and positively associated with milestone and NP status (Bittmann et al., 2022; Bornmann & Tekles, 2021; Wei et al., 2023). 2) Some $DI_1$ variants offer considerable improvements compared to the $DI_1$. The comparative performance of disruption indicators has so far been assessed by five studies. The favourable performances of $DI_5$ in four of these studies (Bittmann et al., 2022; Bornmann et al., 2020a; Bornmann & Tekles, 2021; Deng & Zeng, 2023) are contrasted by only one result that does not confirm the convergent validity of the $DI_5$ (S. Wang et al., 2023). Therefore, we conclude that $DI_5$ shows the most consistently favourable performance in the current literature. The fact that DEP also shows some promising results suggests that indices measuring disruption profit from considering the strength of bibliographic coupling links between the FP and its citing papers. Still, researchers should be aware of the limitations of $DI_5$: $DI_5$ suffers from the inconsistency caused by $N_R$ and its application requires the use of an arbitrary threshold.

## 6.2 Limitations of validation studies

As with any type of research, it should be kept in mind that the studies on the convergent validity of the $DI_1$ and its variants have their own limitations:

(1) Expert judgements and self-assessments used as benchmarks may be flawed in their own way since they may be biased (Bornmann & Daniel, 2009). For example, empirical evidence suggests that expert opinions may be biased against highly novel science (Boudreau et al., 2016).

(2) Since experts have access to bibliometric data, it is possible that they take citations counts into consideration when assigning milestone status to papers. The same applies to the self-assessments of researchers.

(3) If disruption scores are related to citation counts, the results of validation studies could be confounded by citation impact (Bittmann et al., 2022; Bornmann & Tekles, 2021).

(4) The studies relying on the *Faculty Opinions* database use aspects of novelty to test the convergent validity of the $DI_1$ and its variants. Bornmann et al. (2020a, p. 1256) point out that, because novelty and disruption are distinct concepts, there is no way to "completely exclude the possibility that many nondisruptive discoveries are novel".

## 6.3   Best practice guidelines

On the positive side, the current state of research highlights a key strength of the $DI_1$ and its variants. They are highly versatile indices, which provide researchers with a number of options to tackle some of their weaknesses:

(1) Researchers who want to avoid the inconsistency caused by $N_R$ still have a variety of indices to choose from.

(2) Since publications without (indexed) cited references are a major threat to the validity of disruption scores, it seems to become standard practice to calculate disruption scores only for publications with at least a certain number (e.g. 10) of citations and cited references (Bornmann et al., 2020a; Bornmann & Tekles, 2019b; Deng & Zeng, 2023; Ruan et al., 2021; Sheng et al., 2023).

(3) For the same reason, it seems also advisable not to calculate disruption scores for articles from very early publication years.

(4) Because a short citation window may lead to non-reliable results, Bornmann and Tekles (2019a) propose a citation window of at least three years after publication, as it is usually recommended in bibliometrics (van Raan, 2019). However, a time window of three years does not in any way guarantee reliable results for articles that keep on accruing citations long after publication. Since there is no one-size-fits-all approach to citation windows for the $DI_1$ and its variants, researchers who want to work with the indices are encouraged to provide transparent reasons for their choice of the citation window.

## 6.4   Future research

Future research is needed to address four key aspects that have not been sufficiently covered by the current literature:

(1) The concept of disruption requires a precise definition. In the current literature, disruption

is loosely associated with the idea of scientific breakthrough or paradigm shift. However, Wuestman et al. (2020) showed that there are different types of scientific breakthroughs and that breakthroughs should therefore not be treated as a homogenous group. Consequently, there needs to be a discussion about how the concept of disruptive research fits into the typology of scientific breakthroughs.

(2) The current state of research shows that the time-sensitivity of disruption scores is one of the major limitations of the $DI_1$ and its variants. Time-sensitive bias could render historical analysis with the indices challenging (like Park et al., 2023). Current research on time-sensitive biases that affect $DI_1$ (and other variants) is still in a very early (preprint) stage (Bentley et al., 2023; Macher et al., 2023; Petersen et al., 2023) and requires further support by future publications.

(3) The substantial correlation between citation impact and expert assessments of disruptive papers needs to be examined in more detail. The result seems to contradict the central claim of Funk and Owen-Smith (2017) and Wu et al. (2019) that impact measures are not able to identify disruptive papers or patents.

(4) The studies which investigated the relationship between disruption scores and different aspects of novelty lead to seemingly conflicting results. The results by Shibayama and Wang (2020) seem to imply that theoretical (and not methodological) innovations contribute to disruptive research, whereas the findings of Leahey et al. (2023) indicate that new methods (and not new theories) are a key driving force behind disruption in science. Given these inconclusive results, more research on the relationship between disruption scores and specific types of novelty is needed. Such research could not only provide valuable insights into the theoretical and technical properties of disruption metrics but also improve our understanding of scientific innovation.

(5) More research is needed on the lesser known variants of the $DI_1$, since some of them (like $mED_{(rel)}$ and $Orig_{base}$) have been examined so far by only one or very few studies. Future research could also explore how the use of different thresholds affects $DI_l$ (e.g. $I = 3, I = 10$, $I = 20$, etc.). The investigation of the $DI_1$ and the development of variants have led to important insights and recommendations of necessary improvements, but it looks as if this research has still not reached its full potential.

# 7 Conclusions

This review was intended to reveal an overview of the research on the $DI_1$ and its variants. Although these indices have been applied already in science of science studies targeting important science policy questions (e.g., do we have more or less disruptiveness in research over time), we would like to encourage more empirical studies on the reliability, validity and other properties of the different indices. These results are necessary to know significantly more about the indices. Only if the properties of the $DI_1$ and its variants are well known, the empirical studies that investigate science using the indices can be properly designed and interpreted.

## Statements and declarations

Competing interests: The authors have no relevant financial or non-financial interests to disclose.

Funding: Open Access funding enabled and organized by Project DEAL.

Lutz Bornmann is Editorial Board Member of *Scientometrics*.